\documentclass[Preprint]{elsarticle}
\usepackage[utf8]{inputenc}
\usepackage[english]{babel}
\usepackage{amsthm}
\usepackage{blkarray}
\newtheorem{Theorem}{Theorem}[section]

\newdefinition{Definition}{Definition}[section]

\newtheorem{Example}{Example}[section]

\usepackage{float}
\usepackage{soul}
\usepackage{graphicx}
\usepackage{amsmath}
\usepackage{enumerate}
\usepackage{amssymb}
\usepackage{setspace}
\usepackage{tikz}
\usepackage[margin=1in]{geometry}
\usepackage{subcaption}
\usepackage{epstopdf}
\usepackage{lineno}
\usepackage[toc,title,page]{appendix}
\usepackage[font=small,labelfont=bf]{caption}
\definecolor{yellow}{rgb}{1,0.7,0}
\date{}

\graphicspath{{C:/Users/Chris/Documents/University/PhD/LatexDocuments/NodeLevelPaper/Submission/Figures/}{M:/Documents/PhDnew/LatexDocuments/NodeLevelPaper/Submission/Figures/}{C:/Users/hscovert/Documents/LatexDocuments/NodeLevelPaper/Submission/Figures/}{M:/Documents/PhD/LatexDocuments/NodeLevelPaper/Submission/Figures/}}

\begin{document}
\title{Methods for approximating stochastic evolutionary dynamics on graphs}
\author[1]{Christopher E. Overton\corref{cor1}}
\ead{C.Overton@liverpool.ac.uk}
\author[2]{Mark Broom}
\author[3]{Christoforos Hadjichrysanthou}
\author[1]{Kieran J. Sharkey}
\cortext[cor1]{Corresponding author}
\address[1]{Department of Mathematical Sciences, University of Liverpool, Mathematical Sciences Building, Liverpool L69 7ZL, UK.}
\address[2]{Department of Mathematics, City, University of London, Northampton Square, London EC1V 0HB, UK.}
\address[3]{Department of Infectious Disease Epidemiology, School of Public Health, Imperial College London, St Mary's Campus, Norfolk Place, London W2 1PG, UK.}
\begin{abstract}
Population structure can have a significant effect on evolution. For some systems with sufficient symmetry, analytic results can be derived within the mathematical framework of evolutionary graph theory which relate to the outcome of the evolutionary process. However, for more complicated heterogeneous structures, computationally intensive methods are required such as individual-based stochastic simulations. By adapting methods from statistical physics, including moment closure techniques, we first show how to derive existing homogenised pair approximation models and the exact neutral drift model. We then develop node-level approximations to stochastic evolutionary processes on arbitrarily complex structured populations represented by finite graphs, which can capture the different dynamics for individual nodes in the population. Using these approximations, we evaluate the fixation probability of invading mutants for given initial conditions, where the dynamics follow standard evolutionary processes such as the invasion process. Comparisons with the output of stochastic simulations reveal the effectiveness of our approximations in describing the stochastic processes and in predicting the probability of fixation of mutants on a wide range of graphs. Construction of these models facilitates a systematic analysis and is valuable for a greater understanding of the influence of population structure on evolutionary processes.
\end{abstract}
\maketitle 

\section{Introduction}
\label{sec:Intro}
Models of evolutionary dynamics were originally deterministic and assumed well-mixed populations in which every individual of a given type is identical. Stochastic models of these finite well-mixed populations have been studied~\cite{Moran1958}, however real populations are usually characterised by a complicated relationship structure between individuals~\cite{Zhangetal2007}. To account for this, a class of mathematical models known as evolutionary graph theory have been developed which show that the population structure can significantly influence the outcome of evolutionary dynamics~\cite{Liebermanetal2005, TraulsenHauert2010}. In these models, structured populations are represented by finite graphs, where each node represents an individual in the population and relationships between individuals are represented by the edges of the graph. Stochastic evolutionary processes can be considered analytically and precise results can be derived for a number of simple graphs, such as the circle, star and complete graphs~\cite{Broometal2010, BroomRychtar2008,Liebermanetal2005}, mainly due to their symmetry. Analytic approaches for investigating evolutionary dynamics on complex graphs have also been proposed. However, such methods are usually limited by assumptions such as large populations \cite{Nowaketal2010,Ohtsukietal2006} or are specifically designed for investigating evolutionary processes under weak selection~\cite{Allenetal2017, Zhongetal2013}, where the evolutionary game has only a small effect on reproductive success. 

Important quantities of interest such as the exact fixation probability and time can, in principle, be obtained by solving the discrete-time difference equations of the underlying stochastic model~\cite{Hindersinetal2016}, although this is only feasible for very small populations unless there are simplifying symmetries. Individual-based stochastic simulations~\cite{Barbosaetal2010, Maciejewskietal2014} provide numerically accurate representations of the evolutionary process on arbitrary graphs but have limited scope for generating conceptual insights into the dynamics on their own. They can also be computationally expensive on larger graphs, but as a precise representation of the underlying stochastic model, they allow us to evaluate the accuracy of approximate models by comparison. 

Here we develop approximations to the stochastic model by using insights from methods in statistical physics that have also been used extensively for epidemic modelling~\cite{BornGreen1946,KeelingEames2005,Kirkwood1947,PellisHouseKeeling2015,Sharkey2008,Sharkey2015}. Such methods have been applied to develop pair approximations for evolutionary processes on graphs which satisfy the homogeneity assumption that all individuals can be considered identical and interchangable~\cite{Hadjichrysanthouetal2012,HauertSzabo2005,Morita2008,PenaVolken2009,SzaboFath2007}. However, the underlying assumptions linking these models to the underlying stochastic dynamics are not always clear. One contribution of this work is to derive these models explicitly by identifying the required assumptions. The starting point for all of our approximations is to derive an equation to describe the time-evolution of the state of any given individual node. From this equation, various routes to approximation become apparent by applying different assumptions. We then investigate the applicability and accuracy of the resulting approximation methods.

Evolutionary graph theory is traditionally explored as a discrete-time stochastic model. While it is possible to work with these dynamics, it is easier to work with a continuous-time approximation to the process. The continuous-time system is represented by a master equation describing how the probability of being in each system state changes. From the master equation we obtain exact equations (with respect to the continuous-time process) for the probabilities of the states of individual nodes (Theorem~\ref{THM:NodeEquation}). These equations can then be approximated by adopting moment-closure methods. We focus on evaluating the probability that at the end of the evolutionary process, an initial subset of mutants placed on the graph will take over the whole population and `fixate'. Using this continuous-time system is justified because the fixation probability and expected time to fixation are identical to those of the original discrete-time process. Within this framework we study when accurate approximations can be derived.

In Sections~\ref{sec:Dynamics}-\ref{sec:Node} we introduce the stochastic evolutionary dynamics and the master equation, and derive a description of how node-level quantities change in the master equation. We then discuss and develop various techniques that can be used to approximate these systems of equations in Section~\ref{sec:Approx}. Within these approximation frameworks we derive the pair approximation models used in the literature, which we will call the homogenised pair approximation, and the exact neutral drift model, and build new node level approximation methods. In Section~\ref{sec:Results} we demonstrate how the different methods can be used to approximate the dynamics of the original discrete-time process. Section~\ref{sec:Pfix} studies how the methods perform when approximating the fixation probability of a single initial mutant placed on idealised and on complex graphs. Section~\ref{sec:HD} then shows how the methods perform when studying the evolutionary game dynamics in a Hawk-Dove game. In Section~\ref{sec:Discussion} we discuss the results obtained from the methods developed and the insights these can give.

\section{The stochastic model}
\subsection{Stochastic evolutionary dynamics}
\label{sec:Dynamics}

We consider a population whose relationship structure is represented by a strongly connected undirected graph $(V,E)$ where $V=\{1,2,...,N\}$ is the set of nodes and $E$ denotes the set of edges. This can be represented by an adjacency matrix $G$, where $G_{ij}=1$ if $j$ is connected to $i$, and $G_{ij}=0$ otherwise, with $G_{ii}=0$ for all $i\in V$. We consider populations consisting of two types of individuals, type $A$ and type $B$, either of which can be in the role of invading mutant in a resident population. Each node is occupied by either an $A$ or a $B$ individual. Therefore we can let $A_i=1$ if and only if node $i$ is occupied by an $A$ individual and $A_i=0$ otherwise and let $B_i$ denote the same for individuals of type $B$. Since $B_i=1-A_i$, the state of the system can be represented by the values of $A_i$ at any given time. If there exists an edge $(i,j)\in E$ between nodes $i,j \in V$, then the offspring of the individual in node $j$ can replace the individual in node $i$ and vice versa. To study the evolutionary dynamics between these two types of individual we require a measure of fitness. We can describe the fitness payoff received from interactions between individuals by the following payoff matrix:
\begin{equation}
\begin{blockarray}{ccc}
& A & B \\
\begin{block}{c(cc)}
A & a & b \\
B & c & d \\
\end{block}
\end{blockarray},\nonumber
\end{equation}
where an $A$ individual obtains a payoff $a$ when interacting with another $A$ individual and payoff $b$ when interacting with a $B$ individual. Similarly, a $B$ individual obtains payoffs $c$ and $d$ when interacting with an $A$ individual and a $B$ individual respectively.

To define fitness based on the payoff, following similar definitions in the literature~\cite{Hadjichrysanthouetal2011,Liebermanetal2005,Ohtsukietal2006,TraulsenHauert2010,Tayloretal2004}, the fitness of each individual is assumed to be $f=f_{back}+wP$, where $f_{back}$ is the background fitness of all individuals, $P$ is the average payoff received from interactions with neighbours, and $w \in [0,\infty)$ is a parameter which controls the contribution of the game payoff to fitness. 

The fitness of an $A$ individual which occupies node $j$, $f_A^j$, is therefore given by
\begin{equation}
f_A^j=f_{back}+w\frac{a\sum\limits_{i=1}^{N}G_{ij}A_i + b\sum\limits_{i=1}^{N}G_{ij} B_i}{\sum\limits_{i=1}^{N}G_{ij}},
\label{eqn:Afit}
\end{equation}
and similarly the fitness of a $B$ individual occupying node $j$ is given by 
\begin{equation}
f_B^j=f_{back}+w\frac{c\sum\limits_{i=1}^{N}G_{ij}A_i + d\sum\limits_{i=1}^{N}G_{ij} B_i}{\sum\limits_{i=1}^{N}G_{ij}}.
\label{eqn:Bfit}
\end{equation}
In the special case of constant fitness, where the fitness of individuals remains constant independent of the interactions with other individuals, we take the payoff matrix as
\begin{equation}
\begin{blockarray}{ccc}
& A & B \\
\begin{block}{c(cc)}
A & r & r \\
B & 1 & 1 \\
\end{block}
\end{blockarray},
\label{const_fit}
\nonumber
\end{equation}
so that $A$ individuals have relative payoff equal to $r$.

Traditional evolutionary graph theory considers a discrete-time Markovian evolutionary process in which only one event can happen at each time step. When an event occurs, one individual reproduces and a connected individual dies, with the offspring replacing it. We refer to the mechanism by which this takes place as an update mechanism or rule. The probability of a certain event taking place depends upon this update mechanism. Some of the most commonly considered update mechanisms are birth-death with selection on birth (invasion process)~\cite{Liebermanetal2005}, death-birth with selection on birth~\cite{Masuda2009}, birth-death with selection on death~\cite{Antaletal2006} and death-birth with selection on death (voter model)~\cite{Ohtsukietal2006}. The methods developed in this paper will be presented in the general case, and can be applied to any of the above update rules, but we shall focus on the invasion process when generating specific examples. In the invasion process, we select an individual to reproduce in proportion to their fitness (selection on birth) and then the offspring replaces a connected individual selected uniformly at random for death (birth then death).

\subsection{The master equation}
To approximate the discrete-time evolutionary process we first translate the discrete-time system to an approximate continuous-time system. To do this we model each (replacement) event using a Poisson process. The rate at which each event happens is equal to the probability of that event in the discrete-time model. Therefore the total event pressure will be the sum of all such probabilities, which is equal to one, so that the time until the next event follows a Poisson process with rate parameter one. We then determine which event takes place using the relevant probability. Under this continuous-time system the fixation probability and expected time to fixation will be identical to those of the discrete-time system, since we use the same probabilities whenever an event occurs and the expected time between events is constant. This is important because these are the main quantities of interest in evolutionary dynamics.

We will use this system to build approximation methods to study the original discrete-time process. We choose to use continuous-time because it enables us to build a system of ordinary differential equations to approximate the dynamics, which allow us to make use of efficient numerical solvers and enable us to derive some analytic results.

Since this evolutionary process is a continuous-time Markov process, we can construct a master equation to describe the dynamics. Let $S_i=(s_1,s_2,...,s_N)$ be a state of the system, where $i \in\{1,...,2^N\}$ and where $s_j=1$ if node $j$ is a type $A$ individual and $s_j=0$ otherwise. We define $S_1=(0,0,...,0)$ and $S_{2^N}=(1,1,...,1)$ to be the states consisting of only $B$ individuals and only $A$ individuals, respectively. 

We introduce a vector $\textbf{p}(t)$ which represents the probabilities of each system state at time $t$. That is, the $ith$ entry of $\textbf{p}(t)$, $p_i(t)$, is the probability that the system is in state $S_i$ at time $t$. This Markovian evolutionary process has $2^N$ possible states and the transitions between them are governed by a $2^N \times 2^N$ transition rate matrix $R$ whose entries depend upon the graph and update mechanism we consider.

We write the rate of change in the state probabilities using the master equation of the Markov process:
\begin{equation}
\frac{d\textbf{p}}{d{t}}= R \textbf{p}. \label{eqn:MatrixMaster}
\end{equation}
Such an equation can be constructed for any graph under a Markovian update mechanism. The absorbing states correspond to the all type $B$ or all type $A$ states, $S_1$ and $S_{2^N}$, so are given by $p_1$ and $p_{2^N}$.

Since we consider a strongly connected adjacency matrix $G$, provided we have at least one type $A$ and one type $B$ it is possible to get to either of the absorbing states and therefore from any mixed initial condition the system will always end up distributed between these two states. We define the fixation probability $P_{fix}^{A}(S(i))$ of type $A$ from an initial state $S(i)$ to be the probability of being in the all $A$ absorbing state, that is
\begin{equation}
P_{fix}^A(S_i)= \lim_{t \to \infty} (p_{2^N}(t)|p_i(0)=1),
\nonumber
\end{equation}
where $p_i(0)$ is the probability of being in the state $S_i$ at time $t=0$. Similarly we define the fixation probability of type $B$ as
\begin{equation}
P_{fix}^B(S_i)= \lim_{t \to \infty} (p_{1}(t)|p_i(0)=1).
\nonumber
\end{equation}
The computational cost of implementing system~(\ref{eqn:MatrixMaster}) increases exponentially with $N$ \cite{Hindersinetal2016}, and thus the computation of the fixation probability becomes infeasible as the population size increases. Therefore it is of interest to build approximation methods. Pair approximations of the master equation have been developed under the homogeneity assumption that all nodes on the underlying graph are identical and interchangeable~\cite{HauertSzabo2005,SzaboFath2007}, which can give interesting insight into the evolutionary dynamics. However the homogeneity assumptions made in these approximations result in the loss of insight into graph and node-specific dynamics, so we aim to develop approximations of the master equation which can capture this information.

\subsection{Node level equations}
\label{sec:Node}
We approximate the master equation by approximating the dynamics of the state probabilities of individual nodes in the population. This is motivated by approaches in statistical physics and epidemic modelling~\cite{BornGreen1946,Kirkwood1947,Sharkey2008,Sharkey2015}, and first requires exact equations describing how the probability of each node being occupied by a certain type changes with time, which can be derived from the master equation~(\ref{eqn:MatrixMaster}).

\begin{Definition}\label{def:RP} Let $\chi(\Omega_{j \to i}^t|S^t)$ denote the rate at which the individual in node $j$ replaces the individual in node $i$ at time $t$ given that the system is in state $S$ at time $t$; we refer to this as the replacement rate.
\label{def:ReplacementRate}
\end{Definition}

\begin{Definition}
$X_C^t$ denotes the event that the set of nodes $C$ is in state $X$ at time $t$; for example $A_{\{i\}}^t$ is the event that node $i$ is in the type $A$ state at time $t$.
\end{Definition} 

Throughout this paper we shall use the shorthand $B_{\{i\}}^tA_{\{j\}}^tX_C^t$ to represent the intersection of events $B_{\{i\}}^t \cap A_{\{j\}}^t \cap X_C^t$.

\begin{Theorem}[]
Under any Markovian update mechanism, for a structured population represented by the adjacency matrix $G$, the rate of change of the probability that the individual in node $i$ is an $A$ individual is

\begin{align}
\frac{d{P(A_{\{i\}}^{t})}}{d{t}} = &\sum\limits_{j=1}^N\sum\limits_{X_{V\backslash\{i,j\}}}G_{ij}P(B_{\{i\}}^tA_{\{j\}}^tX_{V\backslash\{i,j\}}^t) \chi(\Omega_{j \to i}^t| B_{\{i\}}^tA_{\{j\}}^tX_{V\backslash\{i,j\}}^t )\nonumber \\
-&\sum\limits_{j=1}^N\sum\limits_{X_{V\backslash\{i,j\}}}G_{ij}P(A_{\{i\}}^tB_{\{j\}}^tX_{V\backslash\{i,j\}}^t)\chi(\Omega_{j \to i}^t| A_{\{i\}}^tB_{\{j\}}^tX_{V\backslash\{i,j\}}^t ),\label{eqn:ME}
\end{align}
where the sum over $X_{V\backslash\{i,j\}}$ is over all possible states of the nodes $V\backslash\{i,j\}$.
\label{THM:NodeEquation}
\end{Theorem}
\begin{proof}
See Appendix~\ref{Nodetheorem}.\phantom\qedhere
\end{proof}

This theorem can be applied to any update mechanism by choosing an appropriate definition for the replacement rate, $\chi(\Omega_{j \to i}^t)$, which we shall define for the invasion process as an example.

\begin{Example}[Invasion process]
The invasion process is an adaptation of the Moran process~\cite{Moran1958} to structured populations. Each event is determined by selecting an individual to reproduce with probability proportional to its fitness. It produces an identical offspring which replaces one of the connected individuals which is chosen uniformly at random. Therefore the rate at which the individual in node $j$ replaces the individual in node $i$ at time $t$ under the invasion process rules is given by
\begin{equation}
\chi(\Omega_{j \to i}^t|S) =\frac{f_{j}^t|S}{F^t|S}\frac{1}{k_j},
\label{eqn:SelectionProb}
\end{equation}
where $f_{j}^t$ is the fitness of the individual occupying node $j$ at time $t$, $F^t=\sum\limits_{m=1}^Nf_m^t$ is the total fitness of the population, and $k_j$ denotes the degree of node $j$. Here, the factor $f_{j}^t/F^t$ is the rate at which node $j$ is selected to reproduce, and $1/k_j$ is the probability of replacing the neighbouring individual $i$ which is selected uniformly at random.
\end{Example}

When calculating $\chi(\Omega_{j \to i}^t)$ in Equation~\eqref{eqn:ME}, we will use the following expression for the fitness of the individual at a given node $j$ at time $t$,
\begin{equation}
f_{j}^t=f_{back}+wP( A_{\{j\}}^t ) \frac{a\sum\limits_{i=1}^{N}G_{ij}P( A_{\{i\}}^t ) + b\sum\limits_{i=1}^{N}G_{ij} P( B_{\{i\}}^t ) }{\sum\limits_{i=1}^{N}G_{ij}}+ wP( B_{\{j\}}^t ) \frac{c\sum\limits_{i=1}^{N}G_{ij}P( A_{\{i\}}^t ) + d\sum\limits_{i=1}^{N}G_{ij} P( B_{\{i\}}^t )}{\sum\limits_{i=1}^{N}G_{ij}},
\label{eqn:fitness}
\end{equation}
which is a sum of equations~(\ref{eqn:Afit}) and~(\ref{eqn:Bfit}) weighted by the node probabilities. We use this definition because when we evaluate Equation~(\ref{eqn:fitness}) given that the system is in a particular state $S$, as required by Equation~\eqref{eqn:ME}, the values of $P( A_{\{k\}}^t ) $ and $P( B_{\{k\}}^t )$ are either 1 or 0, which leads to the fitness of node j in that particular system state (Equations (1) and (2)). However, by defining fitness in terms of the node probabilities, this allows us to have a description of fitness which we can approximate (see Sections~\ref{sec:Uncond} and~\ref{sec:Pair}).

\section{Approximating the stochastic model}
\label{sec:Approx}
In other fields, such as epidemiology, the construction of node-level equations such as Equation~(\ref{eqn:ME}) can lead to a hierarchy of moment equations whereby these equations are written in terms of pair probabilities, pairs are written in terms of triples and so on, until the full system state size is reached and the hierarchy is closed. This is useful when we can find appropriate closure approximations to close this hierarchy at a low order. However, we see that such an approach cannot be used here because we condition against the full system state in Equation~(\ref{eqn:ME}) which means that the full system size appears even at the first order. We therefore attempt to find other methods to simplify this system of equations.

In this section we will describe three different techniques to derive approximations for this system. The first technique yields a system of equations which become computationally infeasible in some circumstances, but by applying homogeneity assumptions to the underlying graph, we can derive the existing pair approximation models currently used in the literature~\cite{Hadjichrysanthouetal2012,HauertSzabo2005,Morita2008,PenaVolken2009,SzaboFath2007} (Section~\ref{sec:Derive}). To reduce computation costs, we then develop methods based on restricting the number of states which we condition against in the replacement rate. We first obtain a method whose computational complexity scales linearly with the population size $N$ and, after an appropriate scaling, approximates the fixation probability well on a wide range of graphs (Section~\ref{sec:Uncond}). Then, in Section~\ref{sec:Pair}, we obtain a method which, although it scales with $N^2$, provides a good approximation to the evolutionary dynamics over the whole time series for various graphs, and in particular provides a very accurate approximation to the initial dynamics of the evolutionary process on all graphs.

\subsection{Deriving the homogenised pair approximation model}
\label{sec:Derive}
One way of simplifying (\ref{eqn:ME}) is to assume that the fitness $f_{j}^t$ does not need to be normalised by the total fitness $F^t$ in the replacement rate (e.g. as in Equation~(\ref{eqn:SelectionProb}) for the invasion process). This approximation is justified because it does not change the final value to which the exact node-level equations converge (and therefore the fixation probability), and will only transform the time series until fixation. Making this assumption, the node level equations simplify so that we only sum over the neighbours of the individual that we selected based on fitness. That is, when looking at the event where node $j$ replaces node $i$, if we are selecting on death we need to condition against the state of all neighbours of $i$, and if selecting on birth we need to condition against the state of all neighbours of $j$. As an example, we shall assume here that selection occurs on birth so that we require conditioning on the neighbourhood of node $j$, however we can also make similar arguments when selecting on death. Using $\bar{\chi}$ to represent this modification of $\chi$ in (\ref{eqn:ME}) and $Q$ to represent the new probability distribution with the modified time series we obtain

\begin{align}
\frac{d{Q(A_{\{i\}}^{t})}}{d{t}} = &\sum\limits_{j=1}^N\sum\limits_{X_{\mathcal{N}_{j}\backslash\{i\}}}G_{ij}Q(B_{\{i\}}^tA_{\{j\}}^tX_{\mathcal{N}_{j}\backslash\{i\}}^t) \bar{\chi}(\Omega_{j \to i}^t| B_{\{i\}}^tA_{\{j\}}^tX_{\mathcal{N}_{j}\backslash\{i\}}^t )\nonumber \\
-&\sum\limits_{j=1}^N\sum\limits_{X_{\mathcal{N}_{j}\backslash\{i\}}}G_{ij}Q(A_{\{i\}}^tB_{\{j\}}^tX_{\mathcal{N}_{j}\backslash\{i\}}^t)\bar{\chi}(\Omega_{j \to i}^t| A_{\{i\}}^tB_{\{j\}}^tX_{\mathcal{N}_{j}\backslash\{i\}}^t ),\label{eqn:neighbour}
\end{align}
where $\mathcal{N}_j$ is the neighbourhood of node $j$; i.e. all nodes that are connected to $j$. To solve this system exactly requires the development of equations describing how the probability of each possible neighbourhood of nodes changes. This in turn would lead to a hierarchy of equations which is computationally similar to the master equation. However it is possible to develop approximation methods by assuming independence at the level of lower-order terms, such as individuals or pairs of nodes, and approximating the neighbourhood probabilities as a function of these.

For example, we can make a pair approximation by applying Bayes' Theorem and assuming statistical independence at the level of pairs to rewrite the neighbourhood probability in terms of pair probabilities. Applying Bayes' Theorem to the probabilities on the right hand side of Equation~(\ref{eqn:neighbour}) we get

\begin{align}
\frac{d{Q(A_{\{i\}}^{t})}}{d{t}} = &\sum\limits_{j=1}^N\sum\limits_{X_{\mathcal{N}_{j}\backslash\{i\}}}G_{ij}Q(A_{\{j\}}^t)Q(B_{\{i\}}^tX_{\mathcal{N}_{j}\backslash\{i\}}^t|A_{\{j\}}^t) \bar{\chi}(\Omega_{j \to i}^t| B_{\{i\}}^tA_{\{j\}}^tX_{\mathcal{N}_{j}\backslash\{i\}}^t )\nonumber \\
-&\sum\limits_{j=1}^N\sum\limits_{X_{\mathcal{N}_{j}\backslash\{i\}}}G_{ij}Q(B_{\{j\}}^t)Q(A_{\{i\}}^tX_{\mathcal{N}_{j}\backslash\{i\}}^t|B_{\{j\}}^t)\bar{\chi}(\Omega_{j \to i}^t| A_{\{i\}}^tB_{\{j\}}^tX_{\mathcal{N}_{j}\backslash\{i\}}^t ).\label{eqn:Bayes}
\end{align}
If we assume statistical independence of all nodes in the neighbourhood of $j$, given the state of $j$, we can rewrite the neighbourhood probability $Q(A_{\{j\}}^t)Q(B_{\{i\}}^tX_{\mathcal{N}_{j}\backslash\{i\}}^t|A_{\{j\}}^t) $ as
\begin{equation}
Q(A_{\{j\}}^t)Q(B_{\{i\}}^tX_{\mathcal{N}_{j}\backslash\{i\}}^t|A_{\{j\}}^t) \approx Q(A_{\{j\}}^t)Q(B_{\{i\}}^t|A_{\{j\}}^t)\prod\limits_{l \in {\mathcal{N}_{j}\backslash\{i\}}}Q(X_{\{l\}}^t|A_{\{j\}}^t), \nonumber
\end{equation}
where $X_{\{l\}}^t$ is event where node $l$ is in the same state as it is in the event $X_{\mathcal{N}_{j}\backslash\{i\}}^t$. Substituting this into Equation~(\ref{eqn:Bayes}) gives

\begin{align}
\frac{d{Q(A_{\{i\}}^{t})}}{d{t}} \approx &\sum\limits_{j=1}^N\sum\limits_{X_{\mathcal{N}_{j}\backslash\{i\}}}G_{ij}Q(A_{\{j\}}^t)Q(B_{\{i\}}^t|A_{\{j\}}^t)\prod\limits_{l \in {\mathcal{N}_{j}\backslash\{i\}}}Q(X_{\{l\}}^t|A_{\{j\}}^t) \bar{\chi}(\Omega_{j \to i}^t| B_{\{i\}}^tA_{\{j\}}^tX_{\mathcal{N}_{j}\backslash\{i\}}^t )\nonumber \\
-&\sum\limits_{j=1}^N\sum\limits_{X_{\mathcal{N}_{j}\backslash\{i\}}}G_{ij}Q(B_{\{j\}}^t)Q(A_{\{i\}}|B_{\{j\}}^t)\prod\limits_{l \in {\mathcal{N}_{j}\backslash\{i\}}}Q(X_{\{l\}}^t|B_{\{j\}}^t)\bar{\chi}(\Omega_{j \to i}^t| A_{\{i\}}^tB_{\{j\}}^tX_{\mathcal{N}_{j}\backslash\{i\}}^t ).\nonumber 
\end{align}
Since $Q(B_{\{i\}}^t|A_{\{j\}}^t)=Q(B_{\{i\}}^tA_{\{j\}}^t)/Q(A_{\{j\}}^t)$, in order to evaluate these equations we require additional equations describing how pair probabilities change with time or some appropriate closure of pairs in terms of single node probabilities. From the master equation we can derive exact equations describing pairs. For the probability $P(B_{\{i\}}^tA_{\{j\}}^t)$ we obtain

\begin{align}
\frac{d{P(B_{\{i\}}^tA_{\{j\}}^{t})}}{d{t}} = &\sum\limits_{k=1}^N\sum\limits_{X_{V\backslash\{i,j,k\}}}G_{jk}P( B_{\{i\}}^tB_{\{j\}}^tA_{\{k\}}^tX_{V\backslash\{i,j,k\}}^t ){\chi}(\Omega_{k \to j}^t| B_{\{i\}}^tB_{\{j\}}^tA_{\{k\}}^tX_{V\backslash\{i,j,k\}}^t ) \nonumber \\
-&\sum\limits_{k=1}^N\sum\limits_{X_{V\backslash\{i,j,k\}}}G_{jk}P( B_{\{i\}}^tA_{\{j\}}^tB_{\{k\}}^tX_{V\backslash\{i,j,k\}}^t ){\chi}(\Omega_{k \to j}^t| B_{\{i\}}^tA_{\{j\}}^tB_{\{k\}}^tX_{V\backslash\{i,j,k\}}^t ) \nonumber\\
+ &\sum\limits_{k=1}^N\sum\limits_{X_{V\backslash\{i,j,k\}}}G_{ik}P( B_{\{k\}}^tA_{\{i\}}^tA_{\{j\}}^tX_{V\backslash\{i,j,k\}}^t ){\chi}(\Omega_{k \to i}^t|B_{\{k\}}^tA_{\{i\}}^tA_{\{j\}}^tX_{V\backslash\{i,j,k\}}^t ) \nonumber\\
-&\sum\limits_{k=1}^N\sum\limits_{X_{V\backslash\{i,j,k\}}}G_{ik}P( A_{\{k\}}^tB_{\{i\}}^tA_{\{j\}}^tX_{V\backslash\{i,j,k\}}^t ){\chi}(\Omega_{k \to i}^t| A_{\{k\}}^tB_{\{i\}}^tA_{\{j\}}^tX_{V\backslash\{i,j,k\}}^t ).
\label{eqn:PairExact}
\end{align}
We can now apply the same assumption regarding total fitness that we used for the single node probabilities so that

\begin{align}
\frac{d{Q(B_{\{i\}}^tA_{\{j\}}^{t})}}{d{t}} = &\sum\limits_{k=1}^N\sum\limits_{X_{\mathcal{N}_k\backslash\{i,j\}}}G_{jk}Q( B_{\{i\}}^tB_{\{j\}}^tA_{\{k\}}^tX_{\mathcal{N}_k\backslash\{i,j\}}^t )\bar{\chi}(\Omega_{k \to j}^t| B_{\{i\}}^tB_{\{j\}}^tA_{\{k\}}^tX_{\mathcal{N}_k\backslash\{i,j\}}^t ) \nonumber \\
-&\sum\limits_{k=1}^N\sum\limits_{X_{\mathcal{N}_k\backslash\{i,j\}}}G_{jk}Q( B_{\{i\}}^tA_{\{j\}}^tB_{\{k\}}^tX_{\mathcal{N}_k\backslash\{i,j\}}^t )\bar{\chi}(\Omega_{k \to j}^t| B_{\{i\}}^tA_{\{j\}}^tB_{\{k\}}^tX_{\mathcal{N}_k\backslash\{i,j\}}^t ) \nonumber\\
+ &\sum\limits_{k=1}^N\sum\limits_{X_{\mathcal{N}_k\backslash\{i,j\}}}G_{ik}Q( B_{\{k\}}^tA_{\{i\}}^tA_{\{j\}}^tX_{\mathcal{N}_k\backslash\{i,j\}}^t )\bar{\chi}(\Omega_{k \to i}^t|B_{\{k\}}^tA_{\{i\}}^tA_{\{j\}}^tX_{\mathcal{N}_k\backslash\{i,j\}}^t ) \nonumber\\
-&\sum\limits_{k=1}^N\sum\limits_{X_{\mathcal{N}_k\backslash\{i,j\}}}G_{ik}Q( A_{\{k\}}^tB_{\{i\}}^tA_{\{j\}}^tX_{\mathcal{N}_k\backslash\{i,j\}}^t )\bar{\chi}(\Omega_{k \to i}^t| A_{\{k\}}^tB_{\{i\}}^tA_{\{j\}}^tX_{\mathcal{N}_k\backslash\{i,j\}}^t ).
\label{eqn:PairHomogenised}
\end{align} Applying Bayes' Theorem to the neighbourhood probability $Q( B_{\{i\}}^tB_{\{j\}}^tA_{\{k\}}^tX_{\mathcal{N}_k\backslash\{i,j\}}^t )$ we obtain
\begin{equation}
Q( B_{\{i\}}^tB_{\{j\}}^tA_{\{k\}}^tX_{\mathcal{N}_k\backslash\{i,j\}}^t ) = Q(B_{\{j\}}^tA_{\{k\}}^t)Q( B_{\{i\}}^tX_{\mathcal{N}_k\backslash\{i,j\}}^t|B_{\{j\}}^tA_{\{k\}}^t ) \nonumber
\end{equation}
We can now assume statistical independence of the remaining nodes given the state of $j$ and $k$ so that
\begin{equation}
Q( B_{\{i\}}^tB_{\{j\}}^tA_{\{k\}}^tX_{\mathcal{N}_k\backslash\{i,j\}}^t ) \approx Q(B_{\{j\}}^tA_{\{k\}}^t)Q(B_{\{i\}}^t|B_{\{j\}}^tA_{\{k\}}^t)\prod\limits_{l \in \mathcal{N}_{k}\backslash\{i,j\}}Q(X_{\{l\}}^t|B_{\{j\}}^tA_{\{k\}}^t).\nonumber
\end{equation}
Since we know that node $i$ is connected to node $j$ we can assume that given the state of node $j$, the state of node $i$ is independent of node $k$, and similarly the state of any node in the neighbourhood of $k$ is independent of node $j$, which gives us
\begin{equation}
Q( B_{\{i\}}^tB_{\{j\}}^tA_{\{k\}}^tX_{\mathcal{N}_k\backslash\{i,j\}}^t ) \approx Q(B_{\{j\}}^tA_{\{k\}}^t)Q(B_{\{i\}}^t|B_{\{j\}}^t)\prod\limits_{l \in \mathcal{N}_{k}\backslash\{i,j\}}Q(X_{\{l\}}^t|A_{\{k\}}^t).\nonumber
\end{equation}
Substituting this into Equation~(\ref{eqn:PairHomogenised}) gives

\begin{align}
\frac{d{Q(B_{\{i\}}^tA_{\{j\}}^{t})}}{d{t}} \approx &\sum\limits_{k=1}^N\sum\limits_{X_{\mathcal{N}_{k}\backslash\{i,j\}}}G_{jk}Q(B_{\{j\}}^tA_{\{k\}}^t)Q(B_{\{i\}}^t|B_{\{j\}}^t)\prod\limits_{l \in \mathcal{N}_{k}\backslash\{i,j\}}Q(X_{\{l\}}^t|A_{\{k\}}^t)\bar{\chi}(\Omega_{k \to j}^t| B_{\{i\}}^tB_{\{j\}}^tA_{\{k\}}^tX_{\mathcal{N}_{k}\backslash\{i,j\}}^t ) \nonumber \\
-&\sum\limits_{k=1}^N\sum\limits_{X_{\mathcal{N}_{k}\backslash\{i,j\}}}G_{jk}Q(A_{\{j\}}^tB_{\{k\}}^t)Q(B_{\{i\}}^t|A_{\{j\}}^t)\prod\limits_{l \in \mathcal{N}_{k}\backslash\{i,j\}}Q(X_{\{l\}}^t|B_{\{k\}}^t)\bar{\chi}(\Omega_{k \to j}^t| B_{\{i\}}^tA_{\{j\}}^tB_{\{k\}}^tX_{\mathcal{N}_{k}\backslash\{i,j\}}^t ) \nonumber\\
+ &\sum\limits_{k=1}^N\sum\limits_{X_{\mathcal{N}_{k}\backslash\{i,j\}}}G_{ik}Q(A_{\{i\}}^tB_{\{k\}}^t)Q(A_{\{j\}}^t|A_{\{i\}}^t)\prod\limits_{l \in \mathcal{N}_{k}\backslash\{i,j\}}Q(X_{\{l\}}^t|B_{\{k\}}^t)\bar{\chi}(\Omega_{k \to i}^t| A_{\{i\}}^tA_{\{j\}}^tB_{\{k\}}^tX_{\mathcal{N}_{k}\backslash\{i,j\}}^t ) \nonumber\\
-&\sum\limits_{k=1}^N\sum\limits_{X_{\mathcal{N}_{k}\backslash\{i,j\}}}G_{ik}Q(B_{\{i\}}^tA_{\{k\}}^t)Q(A_{\{j\}}^t|B_{\{i\}}^t)\prod\limits_{l \in \mathcal{N}_{k}\backslash\{i,j\}}Q(X_{\{l\}}^t|A_{\{k\}}^t)\bar{\chi}(\Omega_{k \to i}^t| B_{\{i\}}^tA_{\{j\}}^tA_{\{k\}}^tX_{\mathcal{N}_{k}\backslash\{i,j\}}^t ).\nonumber
\end{align}
While this system is closed, its computational complexity increases exponentially with the maximum node degree of the graph, so it is not numerically feasible for graphs with highly connected nodes. While this could potentially be addressed by introducing approximations for nodes with high degree and this may lead to accurate models, here we continue towards a simplified model. To do this, we follow the same process as in epidemic models and make a homogeneity assumption by assuming that any pair is equally likely to be in any given state \cite{Kissetal2017,Sharkey2008}; i.e. $Q(X_{\{i\}}^t|Y_{\{j\}}^t)=Q(X^t|Y^t)$ for all pairs $(i,j)$. This leads to 

\begin{align}
\frac{d{Q(A_{\{i\}}^{t})}}{d{t}} \approx &\sum\limits_{j=1}^N\sum\limits_{X_{\mathcal{N}_{j}\backslash\{i\}}}G_{ij}Q(A_{\{j\}}^t)Q(B^t|A^t) ^{k_j-n_X}Q(A^t|A^t)^{n_X} \bar{\chi}(\Omega_{j \to i}^t| B_{\{i\}}^tA_{\{j\}}^tX_{\mathcal{N}_{j}\backslash\{i\}}^t )\nonumber \\
-&\sum\limits_{j=1}^N\sum\limits_{X_{\mathcal{N}_{j}\backslash\{i\}}}G_{ij}Q(B_{\{j\}}^t)Q(A^t|B^t)^{n_X+1}Q(B^t|B^t)^{k_j-n_X-1}\bar{\chi}(\Omega_{j \to i}^t| A_{\{i\}}^tB_{\{j\}}^tX_{\mathcal{N}_{j}\backslash\{i\}}^t ),\nonumber 
\end{align}
where $k_j$ is the degree of node $j$ and $n_X$ is the number of type $A$ individuals in state $X_{\mathcal{N}_{j}\backslash\{i\}}$. Since the transition rate only depends on the number of type $A$ and type $B$ individuals in the neighbourhood of node $j$ and not on their positions, the summand on the right hand side is equal for all states $X_{\mathcal{N}_{j}\backslash\{i\}}$ which have the same configuration of $A$ and $B$ individuals. The frequency of a certain neighbourhood state across all possible configurations is given by the binomial coefficient, so that

\begin{align}
\frac{d{Q(A_{\{i\}}^{t})}}{d{t}} \approx &\sum\limits_{j=1}^N\sum\limits_{n=0}^{k_j-1}G_{ij}{{k_j-1}\choose{n}} Q(A_{\{j\}}^t)Q(B^t|A^t) ^{k_j-n}Q(A^t|A^t)^{n} \bar{\chi}(\Omega_{j \to i}^t|n )\nonumber \\
-&\sum\limits_{j=1}^N\sum\limits_{n=0}^{k_j-1}G_{ij}{{k_j-1}\choose{n}} Q(B_{\{j\}}^t)Q(A^t|B^t)^{n+1}Q(B^t|B^t)^{k_j-n-1}\bar{\chi}(\Omega_{j \to i}^t|n)\nonumber,
\end{align}
where $\bar{\chi}(\Omega_{A \to B}^t|n)$ is the rate at which we select one of the type $A$ individuals to reproduce and replace a type $B$, given that there are $n$ type $A$ individuals and $k_j-n$ type $B$ individuals in the neighbourhood of the selected node.

Since we have assumed that any pair is equally likely, this assumption only holds when every node in the graph forms $k$ connections, which are chosen at random. Therefore we require that node $i$ is equally likely to be connected to any other node and all nodes are topologically equivalent, so that the probability that a given node of type $B$ is connected to $x$ type $A$ neighbours is given by a binomial distribution with $n=k$ and $p=Q(A^t|B^t)$. Therefore the probability of an individual being type $A$ changes with rate

\begin{align}
\frac{d{Q(A^{t})}}{d{t}} \approx & \ kQ(A^t|B^t)Q(B^t)\sum\limits_{n=0}^{k-1}{{k-1}\choose{n}}Q(B^t|A^t) ^{k-n}Q(A^t|A^t)^{n} \bar{\chi}(\Omega_{A \to B}^t|n )\nonumber \\
-& \ kQ(B_t|A_t)Q(A^t)\sum\limits_{n=0}^{k-1}{{k-1}\choose{n}}Q(A^t|B^t)^{n+1}Q(B^t|B^t)^{k-n-1}\bar{\chi}(\Omega_{B \to A}^t|n+1).\nonumber
\end{align}
We can also apply these assumptions to the pair-level equations to obtain a closed system of equations which are efficient to solve numerically. The resulting model is equivalent to the model in~\cite{Morita2008}, which was justified by using the assumption that the population occupies a regular graph, such that all individuals have degree $k$, and that all nodes are topologically equivalent, such that every pair of individuals is equally likely to be connected. We have shown that by applying these assumptions to the exact node-level equations (Equation~(\ref{eqn:ME})) we can derive these models. 

Similarly we can obtain a pair approximation model for the dynamics where we select on death by conditioning against the state of the neighbours of node $i$. Applying analogous assumptions to the previous example then leads to the model in~\cite{Hadjichrysanthouetal2012}. These models have been shown to yield interesting qualitative results about the relative strengths of different strategies in evolutionary games on graphs. However, the homogeneity assumptions made result in losing important aspects of the structure, such as how individual nodes in the system can behave differently. In the next sections we will attempt to develop approximation methods which can capture this node-specific information.

As we alluded to earlier, a natural method would be to use Equation~(\ref{eqn:neighbour}) as a basis for this. However, difficulties in implementing this method on general networks as well as the number of equations that result leads us to a different direction for the present work.

\subsection{An unconditioned fitness approximation model}
\label{sec:Uncond}
Here we develop a method which removes the need to include the probability of whole neighbourhoods by removing the conditioning in the replacement rate. This causes the replacement rate to only depend on the marginal probabilities of the state of each node rather than the full system state. This assumption also motivated a model in~\cite{SzaboFath2007} in which the authors construct a population-level approximation describing how the expected number of individuals of each type change with time. Under this assumption, Equation~(\ref{eqn:ME}) becomes

\begin{equation}\frac{d{{P}(A_{\{i\}}^{t})}}{d{t}} \approx \sum\limits_{j=1}^N\sum\limits_{X_{V\backslash\{i,j\}}}G_{ij}{P}( B_{\{i\}}^tA_{\{j\}}^tX_{V\backslash\{i,j\}}^t ) \chi(\Omega_{j \to i}^t)
-\sum\limits_{j=1}^N\sum\limits_{X_{V\backslash\{i,j\}}}G_{ij}{P}( A_{\{i\}}^tB_{\{j\}}^tX_{V\backslash\{i,j\}}^t ) \chi(\Omega_{j \to i}^t).\nonumber \end{equation}
Since $\chi(\Omega_{j \to i}^t)$ is now the same for all system states,
\begin{equation}
\frac{d{{P}(A_{\{i\}}^{t})}}{d{t}} \approx \sum\limits_{j=1}^NG_{ij}{P}( B_{\{i\}}^tA_{\{j\}}^t ) \chi(\Omega_{j \to i}^t)
-\sum\limits_{j=1}^NG_{ij}{P}( A_{\{i\}}^tB_{\{j\}}^t ) \chi(\Omega_{j \to i}^t).
\nonumber
\end{equation}
Adding and subtracting $\sum\limits_{j=1}^NG_{ij} {P}( A_{\{i\}}^t A_{\{j\}}^t ) \chi(\Omega_{j \to i}^t)$ we obtain

\begin{align}
\frac{d{{P}(A_{\{i\}}^{t})}}{d{t}} \approx &\sum\limits_{j=1}^N\left[G_{ij}\bar{P}( B_{\{i\}}^tA_{\{j\}}^t) \chi(\Omega_{j \to i}^t)+G_{ij}{P}( A_{\{i\}}^tA_{\{j\}}^t) \chi(\Omega_{j \to i}^t)\right]\nonumber\\
-&\sum\limits_{j=1}^N\left[G_{ij}{P}( A_{\{i\}}^tB_{\{j\}}^t ) \chi(\Omega_{j \to i}^t)+G_{ij}\bar{P}( A_{\{i\}}^tA_{\{j\}}^t ) \chi(\Omega_{j \to i}^t)\right] \nonumber \\
\approx &\sum\limits_{j=1}^NG_{ij}{P}( A_{\{j\}}^t) \chi(\Omega_{j \to i}^t)
-\sum\limits_{j=1}^NG_{ij}{P}( A_{\{i\}}^t ) \chi(\Omega_{j \to i}^t),
\nonumber
\end{align}
which is a closed set of $N$ equations with at most $N$ summands on the right hand side. Therefore by defining $\bar{P}$ as an approximation to the probability distribution $P$ we obtain the closed system
\begin{equation}
\frac{d{\bar{P}(A_{\{i\}}^{t})}}{d{t}} = \sum\limits_{j=1}^NG_{ij}\bar{P}( A_{\{j\}}^t)  \chi(\Omega_{j \to i}^t)
-\sum\limits_{j=1}^NG_{ij}\bar{P}( A_{\{i\}}^t ) \chi(\Omega_{j \to i}^t),
\label{eqn:Uncond}
\end{equation}
which is easy to solve numerically for an arbitrary graph.

\begin{Example}[Neutral drift]
In the special case of neutral drift, i.e. when all individuals have identical fitness, the unconditioned fitness model gives the exact fixation probability. With the dynamics of the invasion process under neutral drift we obtain $\chi(\Omega_{j \to i}^t)=\frac{1}{Nk_j}$, and therefore Equation~(\ref{eqn:Uncond}) can be written as
\begin{equation}
\frac{d{\bar{P}(A_{\{i\}}^{t})}}{d{t}} = \sum\limits_{j=1}^NG_{ij}\bar{P}( A_{\{j\}}^t) \frac{1}{Nk_j}
-\sum\limits_{j=1}^NG_{ij}\bar{P}( A_{\{i\}}^t ) \frac{1}{Nk_j},
\label{eqn:Neutral}
\nonumber
\end{equation}
which is equivalent to the exact node equation~(\ref{eqn:ME}) for the invasion process under neutral drift~\cite{Shakarianetal2013}. The unconditioned fitness model is also exact for all update mechanisms under neutral drift, but we do not write the equations explicitly here.
\end{Example}

As the population size $N$ increases, the solution to Equation~(\ref{eqn:Uncond}) moves further away from the exact fixation probability obtained either by solving the master equation~(\ref{eqn:MatrixMaster}) or from the output of stochastic simulations. To obtain a reasonable approximation to the fixation probability from a given initial condition we construct a scaling factor for the constant fitness case by comparing the ratio between the solution of Equation~(\ref{eqn:Uncond}) on a complete graph to the exact fixation probability on a complete graph. We choose the complete graph because the exact fixation probability can be calculated analytically in this case. Whilst we consider the constant fitness case, it may also be possible to find a suitable scaling factor in the frequency dependent fitness case, however using a complete graph may no longer be appropriate because the relative strength of different strategies in some games is strongly affected by the average degree of the graph~\cite{Ohtsukietal2006}.

\begin{Example}[Invasion process]
For constant fitness under the dynamics of the invasion process, the exact fixation probability for $m$ initial mutant $A$ individuals on a complete graph is equivalent to the Moran probability~\cite{Liebermanetal2005}:
\begin{equation}
\rho=\frac{1-\frac{1}{r^m}}{1-\frac{1}{r^N}}.
\nonumber
\end{equation}

Since the fixation probability is known, we now need to solve Equation~(\ref{eqn:Uncond}) on the complete graph to derive the ratio between the two. In the constant fitness case this can be done analytically, with the scaling factor for $m$ initial mutants given by

\begin{equation} \frac{\rho}{\lim\limits_{t \to \infty}A_c(t)}=\frac{\frac{1-\frac{1}{r^m}}{1-\frac{1}{r^N}}}{\frac{1}{r-1}\left(-1+\sqrt{1+\frac{m(r^2-1)}{N}}\right)},\label{eqn:Scale}\end{equation}
where $A_c(t) = \frac{1}{N}\sum\limits_{j=1}^N\bar{P}( A_{\{j\}}^t )$. The derivation of this can be found in Appendix~\ref{proof:Scale2}.
\end{Example}

We can now define two methods for predicting the fixation probability under any Markovian update mechanism.
\begin{itemize}
\item{\textbf{Method 1}~(Unconditioned fitness model) Solve Equation~(\ref{eqn:Uncond}) to provide an approximation to the dynamics of the evolutionary process. (MATLAB code for solving the unconditioned fitness model is provided as supplementary material.)}
\item{\textbf{Method 2}~(Scaled unconditioned fitness model) Solve Equation~(\ref{eqn:Uncond}) and then use a scaling factor, the ratio of the exact fixation probability and the solution to Equation~(\ref{eqn:Uncond}) for the complete graph, to provide an approximation to the fixation probability from a given initial condition.}
\end{itemize}
In Section~\ref{sec:Results} we investigate the numerical performance of these two methods. Note that for the purpose of this paper we have found the scaling factor for Method 2 under the invasion process (Equation~(\ref{eqn:Scale})). However, the method can be applied to other update mechanisms, such as death-birth with selection on birth, by finding an appropriate scaling factor, which can be done by solving Equation~(\ref{eqn:Uncond}) (either analytically or numerically) and comparing to the exact fixation probability on the complete graph. For example, see~\cite{HindersinTraulsen2015} for the exact fixation probability on a complete graph under the DB-B dynamics.

\subsection{A contact conditioning approximation model}
\label{sec:Pair}
In Section~\ref{sec:Uncond} we restricted the conditioning so that we only require the marginal probabilities of the individual nodes. However, this removes a significant amount of information from the dynamics. In the evolutionary process, when considering a replacement event the two nodes of most interest are the node selected for birth and the node selected for death. Therefore, here we follow a similar method but retain conditioning on the states of these two key nodes. Since we restrict the conditioning to only the states of the relevant contact, when looking at the term $\chi(\Omega_{j \to i}^t | B_{\{i\}}^tA_{\{j\}}^tX_{V\backslash\{i,j\}}^t)$ in Equation~(\ref{eqn:ME}) we condition only on the states of the nodes $i$ and $j$ and obtain
\begin{equation}
\chi(\Omega_{j \to i}^t | B_{\{i\}}^tA_{\{j\}}^tX_{V\backslash\{i,j\}}^t) \approx \chi(\Omega_{j \to i}^t | B_{\{i\}}^tA_{\{j\}}^t). 
\nonumber
\end{equation}
Under the above condition, Equation~(\ref{eqn:ME}) becomes

\begin{align}
\frac{d{P(A_{\{i\}}^{t})}}{d{t}} \approx &\sum\limits_{j=1}^N\sum\limits_{X_{V\backslash\{i,j\}}}G_{ij}P( B_{\{i\}}^tA_{\{j\}}^tX_{V\backslash\{i,j\}}^t ) \chi(\Omega_{j \to i}^t| B_{\{i\}}^tA_{\{j\}}^t )\nonumber \\
-&\sum\limits_{j=1}^N\sum\limits_{X_{V\backslash\{i,j\}}}G_{ij}P( A_{\{i\}}^tB_{\{j\}}^tX_{V\backslash\{i,j\}}^t ) \chi(\Omega_{j \to i}^t| A_{\{i\}}^tB_{\{j\}}^t ).
\label{eqn:Contact1}
\end{align}
To see the effect of this assumption on the rates, consider $\chi(\Omega_{j \to i}^t |B_{\{i\}}^tA_{\{j\}}^t)$. Here we condition only against node $i$ being in state $B$ and node $j$ being in state $A$ rather than against the entire system state. Consequently in the fitness equation~(\ref{eqn:fitness}) we have $P( B_{\{i\}}^t ) =1$ and $P( A_{\{j\}}^t )=1$ giving
\begin{equation}
f_{j}^t|B_{\{i\}}^tA_{\{j\}}^t=f_{back}+w \frac{bT_{ij} + a\sum\limits_{l \neq i}G_{jl}P( A_{\{l\}}^t ) + b\sum\limits_{l \neq i}G_{jl} P( B_{\{l\}}^t )}{\sum\limits_{l=1}^{N}G_{jl}}.
\nonumber
\end{equation}
In Equation~(\ref{eqn:Contact1}), the chance of selecting node $j$ is now independent of the state $X_{V\backslash\{i,j\}}^t$ of the remaining nodes which enables the equation to be reduced to
\begin{equation}
\frac{d{P(A_{\{i\}}^{t})}}{d{t}} \approx \sum\limits_{j=1}^NG_{ij}P( B_{\{i\}}^tA_{\{j\}}^t ) \chi(\Omega_{j \to i}^t| B_{\{i\}}^tA_{\{j\}}^t )
-\sum\limits_{j=1}^NG_{ij}P( A_{\{i\}}^tB_{\{j\}}^t ) \chi(\Omega_{j \to i}^t| A_{\{i\}}^tB_{\{j\}}^t ).
\label{eqn:ContactIndiv}
\end{equation}
This gives an approximate equation for individuals in terms of pairs. We then need to build equations to describe pair-level probabilities. Similar methodologies have been followed to describe epidemics propagated on networks~\cite{Sharkey2008,Sharkey2015}. 

Applying the same conditioning to the exact pair level equation~(\ref{eqn:PairExact}) we obtain

\begin{align}
\frac{d{P(B_{\{i\}}^tA_{\{j\}}^{t})}}{d{t}} \approx \sum\limits_{k=1}^NG_{jk}P( B_{\{i\}}^tB_{\{j\}}^t{A}_{\{k\}}^t ) \chi(\Omega_{k \to j}^t| B_{\{j\}}^t{A}_{\{k\}}^t )
&-\sum\limits_{k=1}^NG_{jk}P( B_{\{i\}}^tA_{\{j\}}^t{B}_{\{k\}}^t ) \chi(\Omega_{k \to j}^t| A_{\{j\}}^t{B}_{\{k\}}^t ) \nonumber\\
+ \sum\limits_{k=1}^NG_{ik}P( {B}_{\{k\}}^tA_{\{i\}}^tA_{\{j\}}^t ) \chi(\Omega_{k \to i}^t|{B}_{\{k\}}^tA_{\{i\}}^t )
&-\sum\limits_{k=1}^NG_{ik}P( {A}_{\{k\}}^tB_{\{i\}}^tA_{\{j\}}^t ) \chi(\Omega_{k \to i}^t| {A}_{\{k\}}^tB_{\{i\}}^t ).
\label{eqn:ContactPair}
\end{align}
Similar formulae can be constructed for all possible pairs, writing pairs in terms of triples. In a similar way, triples can be written in terms of quads and so on, up to the full system size $N$ which is then closed. Therefore, when using this method we obtain a hierarchy similar to the BBGKY (Bogoliubov–Born–Green–Kirkwood–Yvon) hierarchy~\cite{BornGreen1946,Kirkwood1947} in statistical physics. However, here the hierarchy only represents an approximation to the original dynamics. Solving this system exactly is no simpler than evaluating Equation~(\ref{eqn:MatrixMaster}) since evaluating the hierarchy in full is comparable in numerical complexity, so we wish to find approximation methods to reduce this. 

With this hierarchy, we can apply techniques developed in statistical physics to approximate higher-order terms as functions of lower-order terms. In particular we can close the system of equations (\ref{eqn:ContactIndiv}) and (\ref{eqn:ContactPair}) at the level of pairs by approximating all triples in Equation~(\ref{eqn:ContactPair}) in terms of pair-level and individual-level probabilities. Similar techniques have been applied for many stochastic processes including in epidemiology~\cite{KeelingEames2005,Kissetal2017,Sharkey2008,Sharkey2015} and evolutionary dynamics~\cite{HauertSzabo2005,Ohtsukietal2006,SzaboFath2007} leading to models which can be numerically evaluated.

To close the system, we require a functional form that can approximate triple probabilities in terms of individual and pair probabilities. One method is to approximate a triple $P(A_{\{i\}}^tB_{\{j\}}^tC_{\{k\}}^t)$ as the product of all possible pairs among these nodes divided by the product of all individuals, i.e.

\begin{equation}
P(A_{\{i\}}^tB_{\{j\}}^tC_{\{k\}}^t) \approx \frac{ P( A_{\{i\}}^t B_{\{j\}}^t ) P( B_{\{j\}}^t C_{\{k\}}^t ) P( A_{\{i\}}^t C_{\{k\}}^t )}{P( A_{\{i\}}^t )P( B_{\{j\}}^t ) P( C_{\{k\}}^t )}.
\label{eqn:Kirkwood}
\end{equation}
This closure is commonly attributed to Kirkwood~\cite{Singer2004} because it is derived from the Kirkwood superposition which approximates triples in terms of pairs in thermodynamics~\cite{Kirkwood1935,KirkwoodBoggs1942}. This is often applied to nodes $i,j,k$ that form a 3-cycle in the graph, which we call a `closed triple', although it can be applied to any triplet of nodes. It has been shown that this closure maximises the entropy of these thermodynamic systems~\cite{Singer2004}, and it also ensures that symmetry is preserved across the triplet. This closure has commonly been adapted to probabilistic systems, such as the BBGKY hierarchy~\cite{BornGreen1946,Kirkwood1947} and epidemic modelling~\cite{Keeling1999,Sharkey2008,SharkeyWilkinson2015}. However, the Kirkwood closure for probabilities does not define a probability distribution since we can obtain $P(B_{\{i\}}^tA_{\{j\}}^t)+P(B_{\{i\}}^tB_{\{j\}}^t)\neq P(B_{\{i\}}^t)$, which has been observed numerically~\cite{Rogers2011}. In spite of this it has been shown to yield accurate approximations in these probabilistic systems~\cite{Rogers2011,Sharkey2008,Singer2004}.

Another closure can be obtained by applying Bayes' Theorem and assuming statistical independence across the triple given the state of the central node, in this case node $j$. By applying Bayes' Theorem we have

\begin{equation}
P(A_{\{i\}}^tB_{\{j\}}^tC_{\{k\}}^t) = P(A_{\{i\}}^t|B_{\{j\}}^tC_{\{k\}}^t)P(B_{\{j\}}^tC_{\{k\}}^t),\nonumber
\end{equation}
which, when we assume statistical independence of nodes $i$ and $k$ given $j$, simplifies to

\begin{equation}
P(A_{\{i\}}^tB_{\{j\}}^tC_{\{k\}}^t) \approx P(A_{\{i\}}^t|B_{\{j\}}^t)P(B_{\{j\}}^tC_{\{k\}}^t)=\frac{P(A_{\{i\}}^tB_{\{j\}}^t)P(B_{\{j\}}^tC_{\{k\}}^t)}{P(B_{\{j\}}^t)}.
\label{eqn:Open}
\end{equation}
Typically this closure is applied to nodes on a graph where nodes $i$ and $j$ are connected and nodes $j$ and $k$ are connected but where there is no connection between nodes $i$ and $k$, which we call an `open triple'. However, it could be applied to any triplet of nodes. This closure method is thought to be most accurate on trees~\cite{Kissetal2017,Rogers2011,Sharkey2015}, and has been shown to be exact for such graphs under the SIR epidemic model~\cite{Kissetal2015,Sharkey2015,SharkeyWilkinson2015}.

We can adopt either closure to remove triples and close the system. For example, if we are using the Kirkwood closure to approximate all triples in Equation~(\ref{eqn:ContactPair}) we obtain the system of equations

\begin{align}
\frac{d{\bar{P}(A_{\{i\}}^{t})}}{d{t}} = &\sum\limits_{j=1}^NG_{ij}\bar{P}( B_{\{i\}}^tA_{\{j\}}^t ) \chi(\Omega_{j \to i}^t| B_{\{i\}}^tA_{\{j\}}^t )
-\sum\limits_{j=1}^NG_{ij}\bar{P}( A_{\{i\}}^tB_{\{j\}}^t ) \chi(\Omega_{j \to i}^t| A_{\{i\}}^tB_{\{j\}}^t ). \nonumber \\
\frac{d{\bar{P}(B_{\{i\}}^tA_{\{j\}}^{t})}}{d{t}} = &\sum\limits_{k=1}^NG_{jk}\frac{ \bar{P}( B_{\{i\}}^t B_{\{j\}}^t ) \bar{P}( B_{\{j\}}^t A_{\{k\}}^t ) \bar{P}( B_{\{i\}}^t A_{\{k\}}^t )}{\bar{P}( B_{\{i\}}^t )\bar{P}( B_{\{j\}}^t ) \bar{P}( A_{\{k\}}^t )} \chi(\Omega_{k \to j}^t| B_{\{j\}}^t{A}_{\{k\}}^t )\nonumber \\ 
&-\sum\limits_{k=1}^NG_{jk}\frac{ \bar{P}( B_{\{i\}}^t A_{\{j\}}^t ) \bar{P}( A_{\{j\}}^t B_{\{k\}}^t ) \bar{P}( B_{\{i\}}^t B_{\{k\}}^t )}{\bar{P}( B_{\{i\}}^t )\bar{P}( A_{\{j\}}^t ) \bar{P}( B_{\{k\}}^t )} \chi(\Omega_{k \to j}^t| A_{\{j\}}^t{B}_{\{k\}}^t ) \nonumber\\
&+ \sum\limits_{k=1}^NG_{ik}\frac{ \bar{P}( B_{\{k\}}^t A_{\{i\}}^t ) \bar{P}( A_{\{i\}}^t A_{\{j\}}^t ) \bar{P}( B_{\{k\}}^t A_{\{j\}}^t )}{\bar{P}( B_{\{k\}}^t )\bar{P}( A_{\{i\}}^t ) \bar{P}( A_{\{j\}}^t )} \chi(\Omega_{k \to i}^t|{B}_{\{k\}}^tA_{\{i\}}^t )\nonumber \\ 
&-\sum\limits_{k=1}^NG_{ik}\frac{ \bar{P}( A_{\{k\}}^t B_{\{i\}}^t ) \bar{P}( B_{\{i\}}^t A_{\{j\}}^t ) \bar{P}( A_{\{k\}}^t A_{\{j\}}^t )}{\bar{P}( A_{\{k\}}^t )\bar{P}( B_{\{i\}}^t ) \bar{P}( A_{\{j\}}^t )} \chi(\Omega_{k \to i}^t| {A}_{\{k\}}^tB_{\{i\}}^t ),\nonumber
\label{eqn:ContactPairtriples}
\end{align}
where $\bar{P}$ represents the approximation to the probability distribution $P$. However, note that using this closure for all triples will eventually require equations for every pair of nodes in the system, whether they are connected or not.

It is also useful to use a combination of the two methods whereby the Kirkwood closure~(\ref{eqn:Kirkwood}) is used for closed triples, and~(\ref{eqn:Open}) is used for open triples \cite{Keeling1999,Sharkey2008}. In this work we shall use this combined approach to obtain a closed system. However, we find that unlike in epidemiology, this standard approach does not produce good results. We therefore also try using just the Kirkwood closure because this permits explicit correlations between nodes which are not linked, although as indicated above, this substantially increases computational complexity because the system of equations will scale with $N^2$ rather than the number of connected individuals in the graph.

With the contact conditioning model we define two different methods to approximate the evolutionary dynamics.
\begin{itemize}
\item{\textbf{Method 3}~(Open and closed triples)
Solve Equation~(\ref{eqn:ContactIndiv}) together with equations for pairs by using two different closures for different types of triples. First consider a triple $P( A_{\{i\}}^t B_{\{j\}}^t Z_{\{k\}}^t )$, $Z \in \{A,B\}$, where there is no link between nodes $i$ and $k$. We call this an open triple, and can approximate it as
\begin{equation}
P( A_{\{i\}}^t B_{\{j\}}^t Z_{\{k\}}^t ) \approx \frac{ P( A_{\{i\}}^t B_{\{j\}}^t ) P( B_{\{j\}}^t Z_{\{k\}}^t )}{P( B_{\{j\}}^t )}.
\nonumber
\end{equation}
If there exists a link between nodes $i$ and $k$ we call this a closed triple, and approximate this using the Kirkwood closure,
\begin{equation}
P( A_{\{i\}}^t B_{\{j\}}^t Z_{\{k\}}^t ) \approx\frac{ P( A_{\{i\}}^t B_{\{j\}}^t ) P( B_{\{j\}}^t Z_{\{k\}}^t ) P( A_{\{i\}}^t Z_{\{k\}}^t )}{P( A_{\{i\}}^t )P( B_{\{j\}}^t ) P( Z_{\{k\}}^t )}.
\nonumber
\end{equation}
Using this method it is only necessary to use pairs which have a link between them in the graph, and so it scales with $Nd$, where d is the average degree of the graph.}
\item{\textbf{Method 4}~(Kirkwood closure only)
Solve Equation~(\ref{eqn:ContactIndiv}) together with equations for pairs by using the Kirkwood closure for all triples. That is, we approximate any triple $P( A_{\{i\}}^t B_{\{j\}}^t Z_{\{k\}}^t )$, $Z \in \{A,B\}$ as
\begin{equation}
P( A_{\{i\}}^t B_{\{j\}}^t Z_{\{k\}}^t ) \approx\frac{ P( A_{\{i\}}^t B_{\{j\}}^t ) P( B_{\{j\}}^t Z_{\{k\}}^t ) P( A_{\{i\}}^t Z_{\{k\}}^t )}{P( A_{\{i\}}^t )P( B_{\{j\}}^t ) P( Z_{\{k\}}^t )}.
\nonumber
\end{equation}
This method requires the use of every pair of nodes in the system, not just those which are directly connected, and so scales with $N^2$. (MATLAB code for solving the contact conditioning model is provided as supplementary material.)}
\end{itemize}

\section{Results}
\label{sec:Results}

\subsection{A comparison of the different methods: fixation probabilities for constant fitness}
\label{sec:Pfix}

\begin{figure}
\includegraphics[width=1\textwidth]{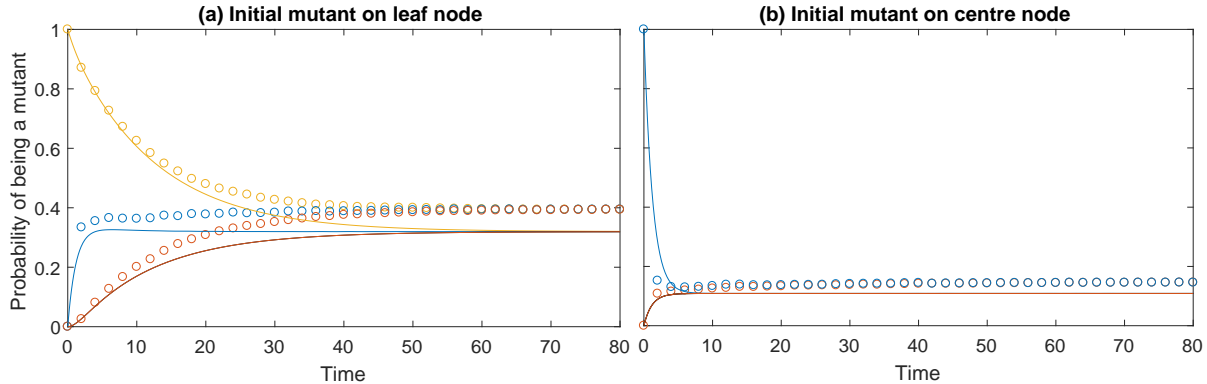}
\caption{Comparison of the marginal probabilities for each node on the graph being a mutant $A$ plotted against time as given by Method 1 (solid lines) versus stochastic simulation of the discrete-time system (circles), when applied to the invasion process on a 4-node star graph. We consider (a) dynamics initiated with a single $A$ individual on a leaf node and (b) dynamics initiated with a single $A$ individual on the central node. Each line represents the marginal probability of a certain node in the graph being occupied by an $A$ individual, the corresponding colours between solid lines and circles represent the same node on the graph. The stochastic process is simulated 10,000 times from the same initial condition until fixation of either the mutant or resident strategy. The probabilities represent, for a given node at a given time, the proportion of simulations for which that node is a mutant. Method 1 is numerically integrated to approximate the probability of each node being a mutant at a given time. This is the constant fitness case where $A$ individuals have fitness $1.2$ and $B$ individuals have fitness $1$.}
\label{fig:LowerBoundCheck}
\end{figure}

Here we investigate the fixation probability of a single initial $A$ individual placed in a given node on the graph under the dynamics of the invasion process. Figure~\ref{fig:LowerBoundCheck} compares Method~1 (unconditioned fitness model) under the invasion process against stochastic simulation on a four-node star graph. On such small graphs, Method~1 appears to provide a reasonable approximation to the expected dynamics and to the fixation probability. However, for such small populations exact solutions are easy to obtain, and hence we want to test larger population sizes. When the population size is increased, this method fails to accurately predict the fixation probability, appearing to tend towards zero with increasing population size (for example, see~Table~\ref{Table:Results1}, where it can be seen that increasing the size from 20 to 35 to 50 moves the solution closer to zero on random graphs). To account for this, we use Method 2~(scaled unconditioned fitness model).

Method 2 represents a scaling of the approximation from Method 1 where the scaling is derived analytically from the fixation probability for a complete graph. Consequently, it makes sense to only consider the approximation of the fixation probability rather than the whole time series. Predictions of the fixation probability of a single $A$ individual when placed on various graphs using the different approximation methods are shown in Tables~\ref{Table:Results1} and~\ref{Table:Results2}. We first observe that the accuracy of the method does not significantly differ for different population sizes, so this overcomes the issue with Method 1. For both the Erd\H{o}s-R{\'e}yni~\cite{ErdosReyni1960} and scale-free random graphs, we start the process in three different initial conditions; a high-degree initial node, a low-degree initial node and an average degree initial node. This is because under the dynamics of the invasion process, a low degree node is known to act as an amplifier of selection and a high degree node is known to act as a suppressor~\cite{Antaletal2006,Shakarianetal2013}, and so we potentially expect different performance of the methods when initiated from nodes of different degree. In the $k$-regular random graph, since all nodes have equal degree, we only consider results for one initial node. In addition to the random graphs (Table~\ref{Table:Results2}), we also investigate a star graph, a square lattice and Zachary’s karate club~\cite{Zachary1977}, which is an example of a real-world network consisting of 34 individuals and average degree of 4.6. On these graphs we initiate the dynamics from a high degree and low degree node. We observe that Method 2 performs best on the $k$-regular random graph and that generally it performs very well on any graph that does not strongly amplify or suppress the average fixation probability compared to the Moran probability, such as the Erd\H{o}s-R{\'e}yni random graph and the square lattice. However on graphs which amplify (or suppress) average fixation probability, such as the scale-free random graph, the approximation becomes less accurate. On the star graph, which significantly amplifies the fixation probability, the approximation is very far from the true value. This is unsurprising because Method 2 is constructed to give the exact fixation probability on complete graphs. For Zachary's karate club, Method 2 provides a reasonable approximation, but does not capture the strong amplifying effect of the low degree node.

\begin{table}

\caption{The fixation probability starting from a single mutant $A$ individual placed on a specific node on single realisations of random graphs. To evaluate the fixation probability using the approximate methods, we solved them until a steady state was reached and calculated the average probability of a node being a mutant (the methods do not always give exactly the same value for each node). We compare this to the fixation probability as calculated by the proportion of 10,000 stochastic simulations in which the type $A$ individuals fixated. Constant fitness is assumed, where $A$ individuals have fitness $1.2$ and $B$ individuals have fitness $1$. All graphs were generated to have an average degree of 5.} 
\begin{center}
\resizebox{0.9\columnwidth}{!}{
\begin{tabular}{| l| c | c |c|c|c|}
\hline
Graph & \multicolumn{5}{|c|}{Fixation probability} \\ \cline{2-6}
& Method 1 & Method 2 & Method 3 & Method 4 & Simulation \\ \hline
20 node Erd\H{o}s-R{\'e}yni - initial degree 10 & 0.0193 & 0.0604 & 1.0000 & 0.0654 & 0.0784\\ \hline
20 node Erd\H{o}s-R{\'e}yni - initial degree 2 & 0.1055 & 0.3301 & 1.0000 & 0.2874 & 0.3098 \\ \hline
20 node Erd\H{o}s-R{\'e}yni - initial degree 5 & 0.0424 & 0.1326 & 1.0000 & 0.1343 & 0.1575 \\ \hline
20 node scale-free - initial degree 10 & 0.0190 & 0.0594 & 1.0000 & 0.0681 & 0.0783 \\ \hline
20 node scale-free - initial degree 2 & 0.0945 & 0.2956 & 1.0000 & 0.3004 & 0.3153\\ \hline
20 node scale-free - initial degree 5 & 0.0475 & 0.1486 & 1.0000 & 0.1490 & 0.1606\\ \hline
20 node $k$-regular & 0.0547 & 0.1711 & 1.0000 & 0.1516 & 0.1722 \\ \hline
35 node Erd\H{o}s-R{\'e}yni - initial degree 10 & 0.0126 & 0.0671 & 1.0000 & 0.0782 & 0.0940\\ \hline
35 node Erd\H{o}s-R{\'e}yni - initial degree 2 & 0.0628 & 0.3346 & 1.0000 & 0.3255 & 0.3191 \\ \hline
35 node Erd\H{o}s-R{\'e}yni - initial degree 5 & 0.0315& 0.1679 & 1.0000 & 0.1572 & 0.1730 \\ \hline
35 node scale-free - initial degree 10 & 0.0089& 0.0474 & 1.0000 & 0.0844 & 0.0724 \\ \hline
35 node scale-free - initial degree 2 & 0.0444 & 0.2366 & 1.0000 & 0.4743 & 0.2929\\ \hline
35 node scale-free - initial degree 5 & 0.0223 & 0.1188 & 1.0000 & 0.1950 & 0.1546 \\ \hline
35 node $k$-regular & 0.0313 & 0.1668 & 1.0000 & 0.1631 & 0.1750 \\ \hline
50 node Erd\H{o}s-R{\'e}yni - initial degree 10 & 0.0083 & 0.0630 & 1.0000 & 0.0787 & 0.0820 \\ \hline
50 node Erd\H{o}s-R{\'e}yni - initial degree 2 & 0.0332 & 0.2521 & 1.0000 & 0.4175 & 0.3060 \\ \hline
50 node Erd\H{o}s-R{\'e}yni - initial degree 5 & 0.0272 & 0.2065 & 1.0000 & 0.2275 & 0.2120 \\ \hline
50 node scale-free - initial degree 10 & 0.0056 & 0.0425 & 1.0000 & 0.0872 & 0.0660 \\ \hline
50 node scale-free - initial degree 2 & 0.0307 & 0.2331 & 1.0000 & 0.3912 & 0.2840 \\ \hline
50 node scale-free - initial degree 5 & 0.0154 & 0.1169 & 1.0000 & 0.1868 & 0.1530 \\ \hline
50 node $k$-regular & 0.0219 & 0.1667 & 1.0000 & 0.1533 & 0.1640 \\ \hline
\end{tabular}
\label{Table:Results1} 
}
\end{center}
\end{table}

\begin{table} 

\caption{The fixation probability starting from a single mutant $A$ individual placed on a specific node on the example graphs. To evaluate the fixation probability using the approximate methods, we solved them until a steady state was reached and calculated the average probability of a node being a mutant (the methods do not always give exactly the same value for each node). We compare this to the fixation probability as calculated by the proportion of 10,000 stochastic simulations in which the type $A$ individuals fixated. Constant fitness is assumed, where $A$ individuals have fitness $1.2$ and $B$ individuals have fitness $1$.}
\begin{center}
\resizebox{0.9\columnwidth}{!}{
\begin{tabular}{| l| c | c | c|c|c|}
\hline
Graph & \multicolumn{5}{|c|}{Fixation probability} \\ \cline{2-6}
& Method 1 & Method 2 & Method 3 & Method 4 & Simulation \\ \hline
20 node star - initial degree 1 & 0.0574 & 0.1796 & 1.0000 & 0.3801 & 0.2895 \\ \hline
20 node star - initial degree 19 & 0.0030 & 0.0094 & 1.0000 & 0.0217 & 0.0184\\ \hline
25 node square lattice - initial degree 2& 0.0662 & 0.2546 & 1.0000 & 0.1532 & 0.2388 \\ \hline
25 node square lattice - initial degree 4& 0.0332 & 0.1277 & 1.0000 & 0.0780 & 0.1444 \\ \hline
34 node Zachary's karate club - initial degree 2 & 0.0482 & 0.2498 & 1.0000 & 0.4285 & 0.3160 \\ \hline
34 node Zachary's karate club - initial degree 16 & 0.0061 & 0.0314 & 1.0000 & 0.0461 & 0.0450 \\ \hline
36 node star - initial degree 1 & 0.0322 & 0.1717 & 1.0000 & 1.0000 & 0.2971\\ \hline
36 node star - initial degree 35 & 0.0009 & 0.0051 & 1.0000 & 0.0209 & 0.0090\\ \hline
36 node square lattice - initial degree 2 & 0.0483 & 0.2646 & 1.0000 & 0.1363 & 0.2462 \\ \hline
36 node square lattice - initial degree 4 & 0.0242 & 0.1326 & 1.0000 & 0.0689 & 0.1385 \\ \hline
49 node star - initial degree 1 & 0.0224 & 0.1697 & 1.0000 & 1.0000 & 0.3070 \\ \hline
49 node star - initial degree 48 & 0.0005& 0.0035 & 1.0000 & 0.0260 & 0.0059 \\ \hline
49 node square lattice - initial degree 2 & 0.0367 & 0.2734 & 1.0000 & 0.1241 & 0.2494 \\ \hline
49 node square lattice - initial degree 4 & 0.0184 & 0.1369 & 1.0000 & 0.0609 & 0.1477 \\ \hline
\end{tabular}
\label{Table:Results2} 
}
\end{center}
\end{table}

In order to improve upon the accuracy of Method 2 we developed the contact conditioning model to retain more information from the system. The contact conditioning model yields a hierarchy which offers no useful reduction in computational complexity, compared to the master equation~(\ref{eqn:ME}). Therefore we developed Method 3 (open and closed triples approximation), analogous to closures used in epidemiology. However, through numerical evaluation we found that this only yields good approximations for simple graphs, such as line graphs and complete graphs for which we have exact analytic results in any case. On other graphs, the fixation probability approximation is equal to 1~(Tables~\ref{Table:Results1} and~\ref{Table:Results2}) for an advantageous mutant of type $A$, and so this method is not particularly informative.

While the specific reason for this convergence to $1$ (or $0$ if the mutant is disadvantageous) is unclear, it seems likely that it is associated with graph-wide correlations caused by having two absorbing states. To address this we developed Method 4 (Kirkwood closure only). Through testing multiple graphs we observe (Tables~\ref{Table:Results1} and \ref{Table:Results2}) that the best results are obtained on Erd\H{o}s-R{\'e}yni and regular random graphs, with some accuracy lost on scale-free random graphs. We observe that on the 20 node star graph, inaccuracies result in a significantly amplified approximation when initiated on the low degree leaf nodes, and for the 35 and 50 node star graphs the approximations initiated on the leaf node are close to $1$. This is potentially due to the time to convergence on large stars being very long, which allows these inaccuracies to compound so that the system converges to this uninformative solution. This failure does not occur on these stars if we reduce the fitness advantage, suggesting that as the size of the star becomes very large the method will only work under weak selection. On random graphs, which do not significantly amplify fixation, this issue is also observed, but only when the fitness advantage of one type is sufficiently high. This issue starts when the fitness advantage is at about $50\%$, below which the solution converges to intermediate values on all random graphs tested. In addition to testing the star graph as an example of an extreme structure, we also tested a square lattice of various sizes, on which we find that Method 4 significantly underestimates the fixation probability. The square lattice is considered as an extreme scenario for this method because it contains many short cycles of order four, for which the correlations are not explicitly captured by the Kirkwood closure, which describes triples. Presenting the star graph and square lattice therefore illustrate the cases where this method is expected to perform least well. Testing Zachary's karate club~\cite{Zachary1977} illustrates how this method might perform on a real world network. On this graph we find that Method 4 provides a reasonable approximation to the fixation probabilities (Table~\ref{Table:Results2}).

\begin{figure}
\includegraphics[width=1\textwidth]{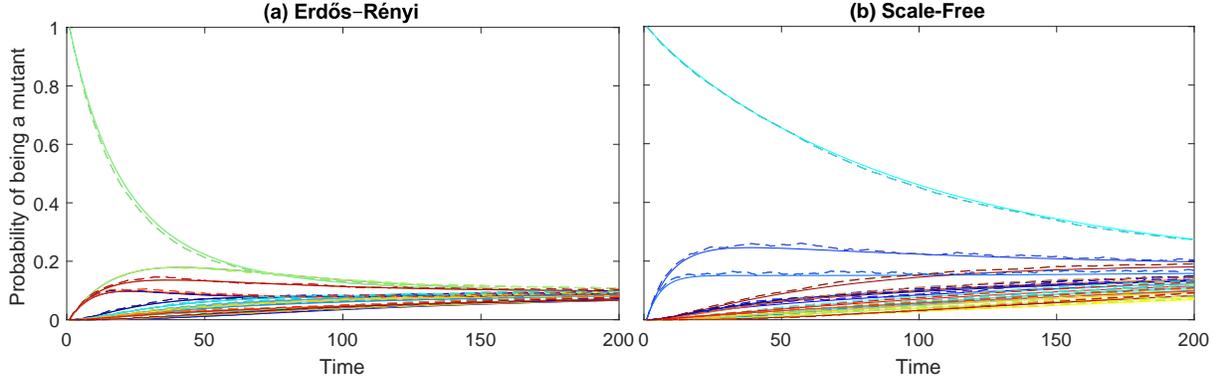}
\caption{Comparison of the early dynamics of the marginal probabilities for each node on the graph being a mutant $A$ plotted against time as given by Method 4 (solid lines) versus stochastic simulation (dashed lines), when applied to the invasion process on (a) an Erd\H{o}s-R{\'e}yni random graph with 20 nodes and average degree of 4 and (b) a scale-free graphwith 20 nodes and average degree 4, both initiated with a single $A$ individual in a chosen node. Each line represents the marginal probability of a certain node in the graph being occupied by an $A$ individual, the corresponding colours between the solid lines and dashed lines represent the same node on the graphs. The discrete-time stochastic process was simulated 10,000 times from the same initial condition, from which we obtained the probability for each node being a mutant at a given time as the proportion of simulations for which that node is a mutant. Method 4 was numerically integrated to approximate the probability of each node being a mutant at a given time. We use a dashed line with interpolation between integer time points for the discrete-time system to enable easier comparison of the dynamics. The game considered is the constant fitness case where the $A$ individuals have fitness $1.2$ and the $B$ individuals have fitness $1$.}
\label{fig:EarlyDynamics}
\end{figure}

We also observed, as shown in Tables~\ref{Table:Results1} and~\ref{Table:Results2}, that Method 4 performs most accurately when initiated on a node with average to high degree. In addition to approximating the fixation probability, Method 4 can be used to approximate the dynamics across the whole time series, and in particular provides a very accurate approximation to the initial dynamics for all graphs tested (see Figure~\ref{fig:EarlyDynamics} for results on two 20 node graphs as an illustration). This accuracy holds even for the large star graphs when initiated on the leaf node, for which the final approximation was close to $1$.

\subsection{The Hawk-Dove game with the contact conditioning model}
\label{sec:HD}
So far, we have considered the constant fitness case. Here we briefly consider the effectiveness of Method 4 when applied to the Hawk-Dove game under the dynamics of the invasion process. Method 2 relies on finding a suitable scaling factor, whilst Methods 1 and 3 were both observed in Section~\ref{sec:Pfix} to yield non informative results on the type of graphs we test here and so we do not investigate these methods in this context. 

\begin{figure}
\includegraphics[width=1\textwidth]{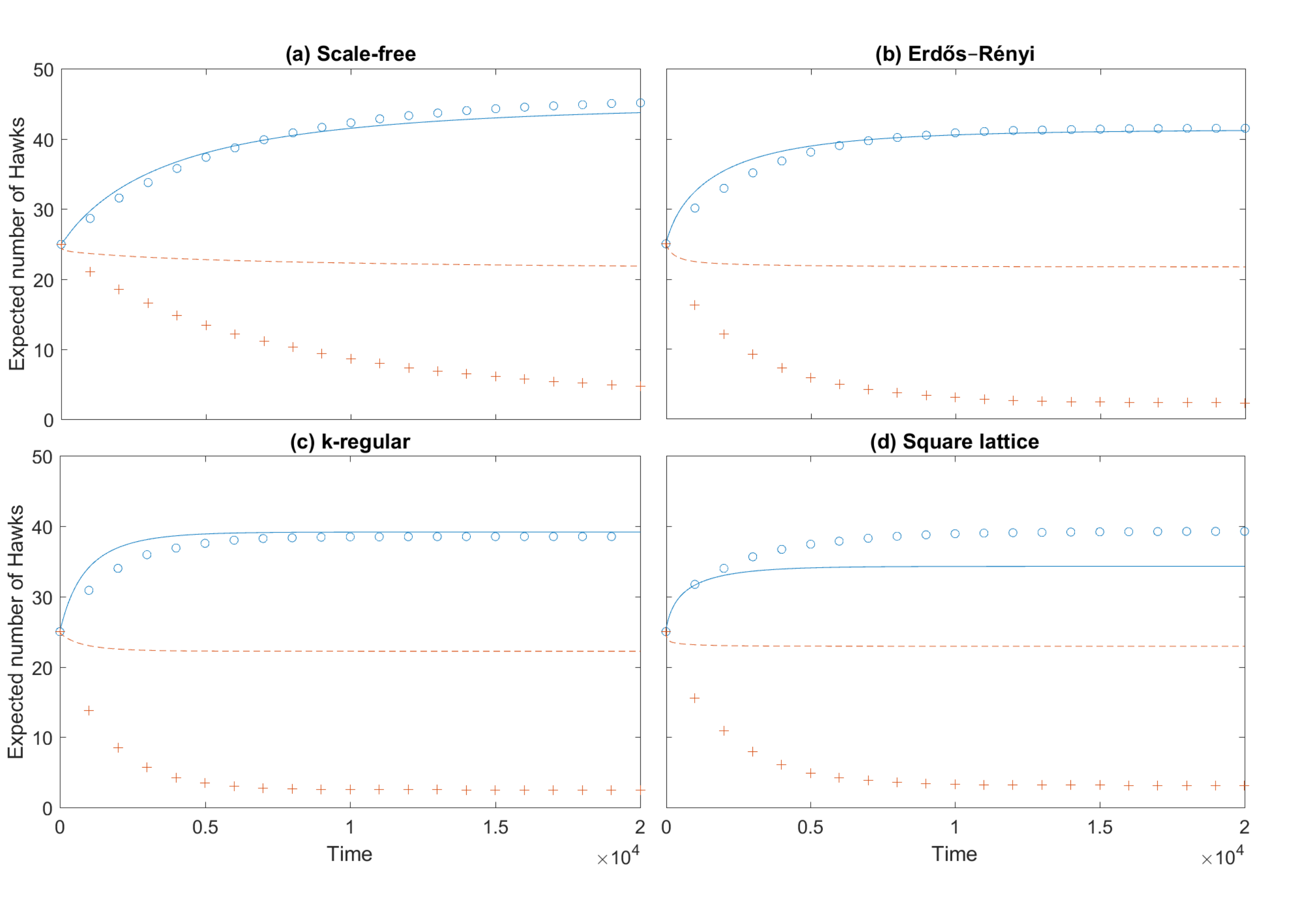}
\caption{Comparison of the expected number of individuals playing the Hawk strategy in a Hawk-Dove game plotted against time as given by Method 4 versus stochastic simulation, when played on (a) a scale-free graph (b) an Erd\H{o}s-R{\'e}yni graph (c) a $k$-regular random graph and (d) a 7 by 7 square lattice. Except for the square lattice, each graph has 50 nodes and an average degree of approximately 4. The solid lines represent the solution of Method 4 and the circles represent stochastic simulations of the discrete-time system, evaluated every 1000 time steps, in the case where $C=1.5$. The dashed lines represent the solution of Method 4 and the crosses represent stochastic simulations of the discrete-time system, evaluated every 1000 time steps, in the case where $C=4$. To generate the stochastic simulation results the discrete-time stochastic process was simulated 10,000 times from the same well mixed initial condition until fixation was reached. By taking the average number of Hawks at each time step we determined the expected number of Hawks at a given time. Method 4 is numerically integrated to give the probability of each node being a Hawk at a given time, from which we obtained the expected number of Hawks by summing over all nodes.}
\label{fig:ApproxResults}
\end{figure}

The Hawk-Dove game~\cite{MaynardSmithPrice1973,MaynardSmith1982} represents a simple model of how animals compete over food, territory and other resources. Animals interact over a resource with either an aggressive or non-aggressive strategy, which we call the Hawk and Dove strategies, respectively. We let the resource yield a payoff $V$ which both players try to obtain. When two Hawks interact, they fight over the resource with one taking the payoff $V$, and the other accruing a cost $C$ from the fight, and therefore the average payoff received by a Hawk interacting with a Hawk is $(V-C)/2$. When a Hawk meets a Dove, the Dove retreats without a fight receiving a payoff $0$, allowing the Hawk to take the whole resource, receiving payoff $V$. If two Doves meet, they either share the resource, or each takes the whole reward without a fight with probability 1/2, so that the average payoff received by a Dove from this interaction is $V/2$. Therefore, in this game the payoff matrix is given by
\begin{equation}
\begin{blockarray}{ccc}
& H & D \\
\begin{block}{c(cc)}
H & (V-C)/2 & V \\
D & 0 & V/2 \\
\end{block}
\end{blockarray}.
\nonumber
\end{equation}
Figure~\ref{fig:ApproxResults} illustrates results from this game on a scale-free graph, an Erd\H{o}s-R{\'e}yni random graph, a $k$-regular random graph and a square lattice. We consider two cases; firstly where the fight cost is low using parameters $f_{back}=2$, $w=1$, $V=1$ and $C=1.5$, and secondly where the fight cost is high using parameters $f_{back}=2$, $w=1$, $V=1$ and $C=4$. In each case we compare the results of Method 4 to stochastic simulation, initiated with a population consisting of half Hawks and half Doves to minimise the chance of early extinction events. We observe that when the cost is low the approximation is reasonable, with all 3 random graphs providing a good approximation, and some accuracy lost on the square lattice. However, as we increase the cost, $C$, we observe that the approximation does not perform well. This is because the contact conditioning assumption seems to amplify the strength of the Hawk strategy, with the rate at which an individual becomes a Hawk under this assumption being greater than it will be in the exact case.

\section{Discussion}
\label{sec:Discussion}
Evolutionary graph theory \cite{Liebermanetal2005} was introduced as a way of adding spatial structure to the stochastic evolutionary dynamics considered by Moran~\cite{Moran1958}. Analytic results on these stochastic dynamics focused on idealised cases of simple graphs \cite{Antaletal2006,Broometal2010}. In order to study arbitrary graphs, methods usually follow certain restrictions, such as focusing on the evolutionary process under weak selection or infinitely large populations~\cite{Allenetal2017,Ohtsukietal2006, Zhongetal2013}. Alternatively, individual-based stochastic simulations give very accurate results but are limited by computational time~\cite{Barbosaetal2010, Maciejewskietal2014}. 

The focus of this work has been the attempt to develop a general method that can approximate the stochastic dynamics on a wide range of graphs by adapting methods from statistical physics and epidemiology. In doing this, we have provided a derivation of existing (homogenised) pair-approximation models from the master equation~\cite{Hadjichrysanthouetal2012,HauertSzabo2005,Morita2008,PenaVolken2009,SzaboFath2007} (Section~\ref{sec:Derive}). Additionally, we also derived an individual-level model which has the neutral drift model~\cite{Shakarianetal2013} as a special case (Section~\ref{sec:Uncond}). 

We start with a representation of the stochastic evolutionary process using a master equation~\cite{Hindersinetal2016}, from which we develop exact equations describing individual node probabilities. We then apply ideas for approximating the master equation based around developing hierarchies of moment equations. Such methods were originally developed in physics~\cite{BornGreen1946,Kirkwood1947} and later used in epidemiology and ecology~\cite{HauertSzabo2005,Keeling1999,PellisHouseKeeling2015,Sharkey2015,SharkeyWilkinson2015}. The key idea behind these techniques is to write deterministic differential equations to describe how the probabilities of the states of individuals and pairs change over time.

We find that a major difference between evolutionary graph theory and other areas in which these methods have been applied is that here, event probabilities depend on the states of all individuals in the population. As a result, we do not obtain a precise BBGKY-like hierarchy, which relies on neighbouring particle-particle interactions. Another difference is that in evolutionary dynamics, we have two absorbing states, which potentially leads to system-wide correlations that cannot be captured on a local level. It is worth noting that some alternative nearest-neighbour interaction evolutionary models, which may yield such a hierarchy directly, have also been considered~\cite{Traulsenetal2005}; however, in this paper we have restricted our attention to the classic evolutionary graph theory dynamics.

In spite of these differences, some progress could be made towards approximating evolutionary dynamics. The first step was to write down equations for the rate of change of the state probabilities for individual nodes (Theorem~\ref{THM:NodeEquation}). This led to equations which required conditioning against the probability of the state of the entire system, and therefore required the development of methods to simplify this. Motivated by an objective of deriving homogenised pair-approximation models used in the literature, our first approach was to modify the replacement rate by removing the normalisation by the total fitness (Section~\ref{sec:Derive}). This has the effect of altering the speed at which events occur but does not alter the final fixation probability. The resulting system of equations describes individual and pair probabilities in terms of the probability of their entire neighbourhoods. This could provide a basis to accurately approximate the fixation probability by finding appropriate moment closures to express the neighbourhoods as functions of individual and pair probabilities. However, this is difficult to implement and the number of equations increases exponentially with the maximum degree of the graph, making it infeasible in general without further approximation. By making further assumptions about the graph such that all  individuals and pairs of a given type are identical and interchangeable, we were able to derive the homogenised pair approximation models~\cite{Hadjichrysanthouetal2012,Morita2008}, which have been shown to give interesting results for various evolutionary games.

To obtain an approximation which is numerically feasible in general, we first ignored any conditioning, similar to a model in~\cite{SzaboFath2007} which uses this assumption to construct a population level approximation. The resulting model (Equation~(\ref{eqn:Uncond})) was found to work well for small graphs and contains the exact neutral drift model~\cite{Shakarianetal2013} as a special case. However, as population size increases, the predictions for the fixation probability of a single mutant individual were observed to tend to zero. By solving this system for the fixation probability on a complete graph, we obtained a scaling factor which enabled this model to give a reasonable prediction of fixation probability from a given initial condition with a single mutant individual on any graph. Due to the construction of this method, it will perform best on graphs which yield average fixation probability close to the Moran probability.

To generate a more accurate model and one which does not require an artificial scaling factor, we investigated models with some level of conditioning (Section~\ref{sec:Pair}). Conditioning against a single node results in the same level of complexity as conditioning against pairs of nodes and so we elected to produce results for the latter. In this case, we conditioned against the pair of nodes directly involved in the replacement event. However, in order to use this model on large graphs, we require the use of moment closure approximations. We found that the standard method used in other areas with different closures for open and closed triples~\cite{Keeling1999, Sharkey2008} was not effective here because while it provides very good results on simple structures, on most graphs it predicts fixation probabilities of either zero or one. It seems likely that this is caused by neglecting important graph-wide correlations across open triples associated with the two absorbing states of the system. 

By using the Kirkwood closure method for all triples, including open ones, we obtained a method which provides informative predictions on the majority of graphs tested. We investigated square lattices and star-type graphs, as these are two extreme population structures which we use as worst case scenarios. The lattice is extreme as moment closure methods do not perform well on such graphs. The star is extreme because this type of graph significantly amplifies fixation probability, which seems to amplify the accumulated error in the approximation methods. For all three types of random graph considered, and Zachary's karate club, this method provides a reasonable approximation to the fixation probability. When the degree of the initial mutant node is not low the approximation can be very accurate. However, if we initiate on a low degree node, the method performs less well, potentially due to such nodes amplifying the fixation probability in the invasion process, again leading to inaccuracies in the solution being amplified. Despite potential inaccuracies in the fixation probability approximation, we observe that this method is particularly accurate for the early-time behaviour of these systems for any graph, and therefore can give interesting insights into this behaviour. The method is computationally feasible for reasonably large $N$, however, the computational complexity scales with $N^2$ rather than with $N$ which is more typical for epidemic models. Nevertheless, this still represents a significant reduction over the master equation which scales with $2^N$. 

The novelty of this work is the adaption of well-established techniques from other fields to the study of evolutionary dynamics at the level of individual nodes. The contribution is two-fold. Firstly we have obtained insight into existing models by deriving them from the master equation. Secondly, the advantage of looking at node-level quantities rather than a homogenised model is that we gain the ability to compare dynamics from different initial conditions on the same graph, which is not present in many other approximation methods. Furthermore, the initial dynamics of Method 4 are very accurate (Figure~\ref{fig:EarlyDynamics}), allowing us to see how the probability of each node being a mutant flows through the population. Although we chose to work in continuous time here and examples study the invasion process, similar methods could be followed directly in discrete-time and the methods are applicable to any Markovian update rule.

\section*{Acknowledgements}
We would like to thank two anonymous reviewers for their valuable comments and suggestions which have contributed to strengthening the manuscript. CO and KS acknowledge support from EPSRC grant (EP/N014499/1). CH and KS acknowledge support from EPSRC grant (EP/K503538/1). MB was supported by funding from the European Union's Horizon 2020 research and innovation programme under the Marie Sklodowska-Curie grant agreement No 690817.

\begingroup
\let\itshape\upshape
\bibliographystyle{plain}
\bibliography{Overton_ApproximatingEvolutionaryDynamics_refs}
\endgroup
\appendix
\renewcommand{\thesection}{\Alph{section}.\arabic{section}}
\setcounter{section}{0}
\begin{appendices}
\section{Proof of Theorem~\ref{THM:NodeEquation}}
\label{Nodetheorem}

\begin{proof}
By total probability rules we have that

\begin{align}
\frac{d{P(A_{\{i\}}^{t})}}{d{t}}= \frac{d\left[{\sum\limits_{X_{V\backslash\{i\}}}P(A_{\{i\}}^tX^t_{V\backslash\{i\}})}\right]}{dt} =\sum\limits_{X_{V\backslash\{i\}}}\frac{d{P(A_{\{i\}}^tX^t_{V\backslash\{i\}})}}{dt},
\label{eqn:Events}
\end{align}
where $X_{V\backslash\{i\}}$ is the state of the nodes in the system not including $i$.

Consider a set state $X_{V\backslash\{i\}}$ of the remaining nodes. The rate of change in the full system state probability $P(A_{\{i\}}^tX^t_{V\backslash\{i\}})$ is given by

\begin{align}
\frac{d{P(A_{\{i\}}^tX^t_{V\backslash\{i\}})}}{dt}=&\sum\limits_{Y_{V\backslash\{i\}}}P(A_{\{i\}}^tY^t_{V\backslash\{i\}})\chi(A_{\{i\}}^tY^t_{V\backslash\{i\}} \to A_{\{i\}}^tX^t_{V\backslash\{i\}})\nonumber \\ \nonumber
&+P(B_{\{i\}}^tX^t_{V\backslash\{i\}})\chi(B_{\{i\}}^tX^t_{V\backslash\{i\}} \to A_{\{i\}}^tX^t_{V\backslash\{i\}}) \\ \nonumber
&-\sum\limits_{Y_{V\backslash\{i\}}}P(A_{\{i\}}^tX^t_{V\backslash\{i\}})\chi(A_{\{i\}}^tX^t_{V\backslash\{i\}} \to A_{\{i\}}^tY^t_{V\backslash\{i\}}) \\ 
&-P(A_{\{i\}}^tX^t_{V\backslash\{i\}})\chi(A_{\{i\}}^tX^t_{V\backslash\{i\}} \to B_{\{i\}}^tX^t_{V\backslash\{i\}}), 
\label{eqn:StateRate}
\end{align}
where $\chi(A_{\{i\}}^tX^t_{V\backslash\{i\}} \to B_{\{i\}}^tX^t_{V\backslash\{i\}}) $ is the rate at which the system moves from state $A_{\{i\}}^tX^t_{V\backslash\{i\}}$ to state $B_{\{i\}}^tX^t_{V\backslash\{i\}}$. 

Consider the terms which involve changing the state of the individual in node $i$ in Equation~(\ref{eqn:StateRate}), by expanding the rate into the sum of separate event rates we obtain

\begin{align}
P(B_{\{i\}}^tX^t_{V\backslash\{i\}})\chi(B_{\{i\}}^tX^t_{V\backslash\{i\}} \to A_{\{i\}}^tX^t_{V\backslash\{i\}})=P(B_{\{i\}}^tX^t_{V\backslash\{i\}})\sum\limits_{j=1}^NG_{ij}\chi(\Omega^t_{j \to i}|B_{\{i\}}^tX^t_{V\backslash\{i\}})\textbf{1}_{(A_{\{j\}}^t \in X^t_{V\backslash\{i\}})}, \nonumber
\end{align}
and

\begin{align}
P(A_{\{i\}}^tX^t_{V\backslash\{i\}})\chi(A_{\{i\}}^tX^t_{V\backslash\{i\}} \to B_{\{i\}}^tX^t_{V\backslash\{i\}})=P(A_{\{i\}}^tX^t_{V\backslash\{i\}})\sum\limits_{j=1}^NG_{ij}\chi(\Omega^t_{j \to i}|A_{\{i\}}^tX^t_{V\backslash\{i\}})\textbf{1}_{(B_{\{j\}}^t \in X^t_{V\backslash\{i\}})}, \nonumber
\end{align}
where $\textbf{1}_{(B_{\{j\}}^t \in X^t_{V\backslash\{i\}})}$ is an indicator function on the event $B_{\{j\}}^t$ being part the event $X^t_{V\backslash\{i\}}$. That is, the state of node $j$ in the state $X$ is type $B$. The $\chi(\Omega^t_{j \to i}|A_{\{i\}}^tX^t_{V\backslash\{i\}})$ term is the rate at which the individual in node $j$ replaces the individual in node $i$, given that the system is in state $A_{\{i\}}^tX^t_{V\backslash\{i\}}$, as defined in Definition~\ref{def:ReplacementRate}. Rearranging these and substituting into Equation~(\ref{eqn:StateRate}) gives

\begin{align}
\frac{d{P(A_{\{i\}}^tX^t_{V\backslash\{i\}})}}{dt}=&\sum\limits_{j=1}^NG_{ij}P(B_{\{i\}}^tX^t_{V\backslash\{i\}})\chi(\Omega^t_{j \to i}|B_{\{i\}}^tX^t_{V\backslash\{i\}})\textbf{1}_{(A_{\{j\}}^t \in X^t_{V\backslash\{i\}})}\nonumber \\ \nonumber
&-\sum\limits_{j=1}^NG_{ij}P(A_{\{i\}}^tX^t_{V\backslash\{i\}})\chi(\Omega^t_{j \to i}|A_{\{i\}}^tX^t_{V\backslash\{i\}})\textbf{1}_{(B_{\{j\}}^t \in X^t_{V\backslash\{i\}})} \\ \nonumber
&+\sum\limits_{Y_{V\backslash\{i\}}}P(A_{\{i\}}^tY^t_{V\backslash\{i\}})\chi(A_{\{i\}}^tY^t_{V\backslash\{i\}} \to A_{\{i\}}^tX^t_{V\backslash\{i\}}) \\ \nonumber
&-\sum\limits_{Y_{V\backslash\{i\}}}P(A_{\{i\}}^tX^t_{V\backslash\{i\}})\chi(A_{\{i\}}^tX^t_{V\backslash\{i\}} \to A_{\{i\}}^tY^t_{V\backslash\{i\}}).
\end{align}
By substituting this into Equation~(\ref{eqn:Events}) we obtain

\begin{align}
\frac{d{P(A_{\{i\}}^{t})}}{d{t}}=&\sum\limits_{X_{V\backslash\{i\}}}\sum\limits_{j=1}^NG_{ij}P(B_{\{i\}}^tX^t_{V\backslash\{i\}})\chi(\Omega^t_{j \to i}|B_{\{i\}}^tX^t_{V\backslash\{i\}})\textbf{1}_{(A_{\{j\}}^t \in X^t_{V\backslash\{i\}})}\nonumber \\ \nonumber
&-\sum\limits_{X_{V\backslash\{i\}}}\sum\limits_{j=1}^NG_{ij}P(A_{\{i\}}^tX^t_{V\backslash\{i\}})\chi(\Omega^t_{j \to i}|A_{\{i\}}^tX^t_{V\backslash\{i\}})\textbf{1}_{(B_{\{j\}}^t \in X^t_{V\backslash\{i\}})} \\ \nonumber
&+\sum\limits_{X_{V\backslash\{i\}}}\sum\limits_{Y_{V\backslash\{i\}}}P(A_{\{i\}}^tY^t_{V\backslash\{i\}})\chi(A_{\{i\}}^tY^t_{V\backslash\{i\}} \to A_{\{i\}}^tX^t_{V\backslash\{i\}}) \\ \nonumber
&-\sum\limits_{X_{V\backslash\{i\}}}\sum\limits_{Y_{V\backslash\{i\}}}P(A_{\{i\}}^tX^t_{V\backslash\{i\}})\chi(A_{\{i\}}^tX^t_{V\backslash\{i\}} \to A_{\{i\}}^tY^t_{V\backslash\{i\}}).
\end{align}
Clearly the last two sums cancel, so we can simplify this to

\begin{align}
\frac{d{P(A_{\{i\}}^{t})}}{d{t}}=&\sum\limits_{j=1}^N\sum\limits_{X_{V\backslash\{i,j\}}}G_{ij}P(B_{\{i\}}^tA_{\{j\}}^tX^t_{V\backslash\{i,j\}})\chi(\Omega^t_{j \to i}|B_{\{i\}}^tA_{\{j\}}^tX^t_{V\backslash\{i,j\}}) \nonumber \\ \nonumber
&-\sum\limits_{j=1}^N\sum\limits_{X_{V\backslash\{i,j\}}}G_{ij}P(A_{\{i\}}^tB_{\{j\}}^tX^t_{V\backslash\{i,j\}})\chi(\Omega^t_{j \to i}|A_{\{i\}}^tB_{\{j\}}^tX^t_{V\backslash\{i,j\}}), 
\end{align}
as required.
\label{proof}
\end{proof}

\section{Derivation of the scaling factor (Equation~\ref{eqn:Scale})}
\begin{proof}
Consider a system with rate of change given by
\[\frac{d\bar{P}({{A}}_{\{i\}}^t)}{dt}= \sum\limits_{j=1}^N G_{ij}\bar{P}( A_{\{j\}}^t )\chi(\Omega_{j \to i}^t) - \sum\limits_{j=1}^N G_{ij}\bar{P}(A_{\{i\}}^t )\chi(\Omega_{j \to i}^t).\]
Since we are interested in the complete graph, we have that $G_{ij}=1$ for $j\neq i$, and $G_{i,i}=0$. Let $A_c$ denote the average probability that a node is of type $A$ on the complete graph at time $t$. That is
\[A_c(t) = \frac{1}{N}\sum\limits_{j=1}^N\bar{P}( A_{\{j\}}^t ) = \frac{S}{N}.\]
Since we are considering constant fitness we have
\[\chi(\Omega_{j \to i}^t)= \frac{\bar{P}( A_{\{j\}}^t ) (r-1) +1}{\sum\limits_{k=1}^N\bar{P}( {A}_{\{k\}}^t ) (r-1) +1}=
\frac{\bar{P}( A_{\{j\}}^t) (r-1) +1}{N+(r-1)S},\]
which gives us
\[\frac{dS}{dt} = \sum\limits_{i=1}^N\frac{d\bar{P}({{A}}_{\{i\}}^t)}{dt}= \frac{\sum\limits_{i,j=1}^N(\bar{P}(A_{\{j\}}^t) - \bar{P}(A_{\{i\}}^t ))(\bar{P}( A_{\{j\}}^t )(r-1)+1)}{N+(r-1)S}.\]
Writing $G$ = $\sum\limits_{i,j=1}^N(\bar{P}( A_{\{j\}}^t) -\bar{P}(A_{\{i\}}^t )) \bar{P}(A_{\{j\}}^t )$, and $H$ = $\sum\limits_{i,j=1}^N(\bar{P}( A_{\{j\}}^t) -\bar{P}(A_{\{i\}}^t )) $ we have
\[\frac{dS}{dt} = \frac{(r-1)G+H}{N+(r-1)S}.\]
Clearly $H=0$, so we obtain

\[\frac{dS}{dt} = \frac{(r-1)G}{N+(r-1)S}.\]
Note that $\sum\limits_{i,j=1}^N (\bar{P}( A_{\{j\}}^t) - \bar{P}(A_{\{i\}}^t ))^2= \sum\limits_{i,j=1}^N \bar{P}( A_{\{j\}}^t )^2 + \bar{P}( A_{\{i\}}^t )^2 -2\bar{P}( A_{\{j\}}^t ) \bar{P}( A_{\{i\}}^t ) = 2G$, so that
\[\frac{dG}{dt}=\frac{1}{2}\frac{d}{dt}\big(\sum\limits_{i,j=1}^N(\bar{P}( A_{\{j\}}^t) - \bar{P}(A_{\{i\}}^t))^2 \big) = \sum\limits_{i,j=1}^N(\bar{P}( A_{\{j\}}^t) - \bar{P}(A_{\{i\}}^t )) \frac{d(\bar{P}( A_{\{j\}}^t) - \bar{P}(A_{\{i\}}^t ))}{dt}.\]
Considering the last term on the right hand side we have

\begin{align*} \frac{d(\bar{P}( A_{\{i\}}^t) - \bar{P}({A}_{\{j\}}^t ))}{dt}&= \frac{1}{N+(r-1)S}\sum\limits_{k=1}^N \big( \bar{P}( A_{\{k\}}^t ) (\bar{P}( A_k)^t - \bar{P}(A_{\{i\}}^t )) + \bar{P}( A_{\{k\}}^t ) (\bar{P}( {A}_{\{j\}}^t) -\bar{P}(A_{\{k\}}^t )) \big)(r-1) \\
&\hspace{3cm}+ (\bar{P}( A_{\{k\}}^t) - \bar{P}(A_{\{i\}}^t )) + (\bar{P}( {A}_{\{j\}}^t) -\bar{P}(A_{\{k\}}^t )) \\
&=\frac{1}{N+(r-1)S}\sum\limits_{k=1}\bar{P}( A_{\{k\}}^t ) (\bar{P}( {A}_{\{j\}}^t) -\bar{P}(A_{\{i\}}^t) )(r-1)+ (\bar{P}( {A}_{\{j\}}^t) -\bar{P}(A_{\{i\}}^t )) \\
&=\frac{(\bar{P}( {A}_{\{j\}}^t) - \bar{P}(A_{\{i\}}^t ))\big((r-1)S+N \big)}{N+(r-1)S} \\
&= -(\bar{P}( A_{\{i\}}^t) - \bar{P}({A}_{\{j\}}^t )).
\end{align*}
Thus,
\[ \frac{dG}{dt} = \sum\limits_{i,j=1}^N (\bar{P}( A_{\{j\}}^t) - \bar{P}(A_{\{i\}}^t ))^2 = -2G \implies G=Ae^{-2t} = (N-m)me^{-2t},\] since $G(0)=(N-m)m$.
Therefore we have
\[\frac{dS}{dt} = \frac{(r-1)(N-m)me^{-2t}}{N+(r-1)S}\]
\begin{align*}
\Rightarrow NS + \frac{r-1}{2}S^2 &= -\frac{1}{2}(r-1)(N-m)me^{-2t} + C.
\end{align*}
At $t=0$ we have $ S=\sum \bar{P}( A_{\{j\}}^t ) = m$, which gives
\[C= Nm +\big(\frac{r-1}{2}\big)Nm = Nm\big( \frac{r+1}{2}\big),\]
and so we can solve to obtain
\[S= \frac{\big(-N \pm \sqrt{N^2 + 4\frac{r-1}{2}\big(Nm\frac{r+1}{2} - (N-m)m\frac{r-1}{2}e^{-2t}\big)}\big)}{r-1}.\]
Only the positive root makes sense, so we obtain
\[A_c = \frac{1}{r-1}\big(-1 + \sqrt{1 + \frac{m(r^2-1)}{N} - (r-1)^2\frac{(N-m)m}{N^{2}}e^{-2t}}\big).\]
Thus, we have $\lim\limits_{t \to \infty} A_c(t) = \frac{1}{r-1}\big(-1 + \sqrt{1 + \frac{m(r^2-1)}{N}}\big)$.
\label{proof:Scale2}
\end{proof}
\end{appendices}

\end{document}